\documentclass[10pt,reqno]{amsart}
\usepackage{graphicx,multicol}
\usepackage{multirow} 
\usepackage{indentfirst,csquotes}
 \usepackage{array} 

\usepackage[caption=false]{subfig}
\usepackage{amssymb,amsthm,amsmath}
\usepackage{xcolor,paralist,hyperref,fancyhdr,etoolbox}

\usepackage[round]{natbib} 

\usepackage[utf8]{inputenc}
\numberwithin{equation}{section}

\hypersetup{ colorlinks=true, linkcolor=black, filecolor=black, urlcolor=black,citecolor = blue }

\usepackage{lipsum}
\usepackage{multicol}

\usepackage{listings}
\begin{document}


\title[]{Semi-parametric least-area linear-circular regression through M\"{o}bius transformation} 

\maketitle
\begin{center}
\author{Surojit Biswas} \\ 
\email{surojit23$@$iitkgp.ac.in}\\ and\\
\author{Buddhananda Banerjee\footnotetext{corresponding author\/: bbanerjee@maths.iitkgp.ac.in}}\\
\email{bbanerjee@maths.iitkgp.ac.in}\\
\vspace{0.3cm}
\address{Department of Mathematics\\  IIT Kharagpur, India-$721302$}

\end{center}

\let\thefootnote\relax

\begin{abstract}
This paper introduces a novel regression model designed for angular response variables with linear predictors, utilizing a generalized Möbius transformation to define the regression curve. By mapping the real axis to the circle, the model effectively captures the relationship between linear and angular components. A key innovation is the introduction of an area-based loss function, inspired by the geometry of a curved torus, for efficient parameter estimation. The semi-parametric nature of the model eliminates the need for specific distributional assumptions about the angular error, enhancing its versatility. Extensive simulation studies, incorporating von Mises and wrapped Cauchy distributions, highlight the robustness of the framework. The model’s practical utility is demonstrated through real-world data analysis of Bitcoin and Ethereum, showcasing its ability to derive meaningful insights from complex data structures.

\end{abstract} 

\newpage
\tableofcontents{}
\bigskip
\newpage
\setlength{\columnsep}{1.5cm} 

\section{Introduction} 
In statistics, the analysis of directional data is an area that addresses measurements of angular random variables which makes it distinctly different from standard linear data due to its periodic nature. This type of data are commonly encountered in different fields such as, angles of astigmatism in ophthalmology \cite[see][]{jha2018circular}, the timing of the highest or lowest price of stocks  in finance, Agency and Communion data in educational psychology \cite[][]{cremers2020regression}, wind and wave directions in meteorology \cite[][]{biswas2024angular},  protein folding problem in bio-informatics \cite[][]{boomsma2008generative}, etc.  A particularly interesting extension of directional data analysis involves angular (circular) component along with a linear covariate, creating a framework suited to analyzes variables that jointly varies.  For example, in finance, the time of occurrence of the highest or lowest price of a stock is represented as angular variable, while the price itself is treated as a linear variable. Similarly, in environmental science, wind measurements often include both the direction (angular) and the speed (linear).

Linear-circular  regression is a statistical approach for analyzing the relationship between a circular (angular) variable and a linear variable. In this framework, the response variable is angular, while the predictor variable is linear. The periodic nature of the circular data, typically contained within a $360^\circ$ or \([0, 2\pi]\) range, introduces unique challenges because of its topology which is not present in  linear regression.
A common approach is to model the circular outcome as a function of the linear predictor, often by expressing the angular component in terms of sine and cosine functions or tangent functions (\citealp[see,][]{fisher1992regression}) to handle periodicity. By effectively linking circular and linear components, linear-circular regression offers valuable insights into relationships that conventional linear regression methods can not provide.

A M\"{o}bius transformation, also known as a linear fractional transformation in complex analysis is a map that projects complex plane, $\mathbb{C}$-including infinity  ($z$-plane), onto itself ($w$-plane) in a way that preserves the general structure of lines and circles. The transformation is defined by the function:
\[
f(z) = \frac{az + b}{cz + d}
\]
where \( z \) is a complex variable, and \( a \), \( b \), \( c \), and \( d \) are complex constants satisfying \( ad - bc \neq 0 \). 
M\"{o}bius transformations possess several remarkable properties such as angle preservation (conformability), circles to lines and/or vice versa.
In a seminal paper, \cite{mccullagh1996mobius}  explored the use of M\"{o}bius transformations in directional statistics to investigate the connection between the real Cauchy distribution and the wrapped Cauchy distribution.

Recently, \cite{jha2018circular} utilized M\"{o}bius transformation as a link function for circular-circular regression, especially on zero-spike data commonly encountered in ophthalmology research. Earlier applications of M\"{o}bius transformations in circular-circular regression models can be found in the works of \cite{}  \cite{downs2002circular}, \cite{downs2003spherical}, and \cite{kato2008circular}. These developments have greatly expanded the scope and utility of M\"{o}bius transformations in regression analysis. At the same time the conditional expectation-based circular-circular regression is proposed by \cite{banerjee2024intrinsic} and linear-circular regression is proposed by \cite{cremers2020regression}.

Here, a new area-based measurement between two angles $\phi, \theta \in \mathbb{S}_1$ is briefly discussed. The rest of our work will be based on the curved torus defined by the parametric equation 
\begin{equation}
  X(\phi,\theta)=\{  (R+r\cos{\theta})\cos{\phi}, (R+r\cos{\theta})\sin{\phi}, r\sin{\theta} \}\subset \mathbb{R}^3, 
  \label{torus para equn}
\end{equation}
with the parameter space $\{ 
 (\phi,\theta):0<\phi,\theta<2\pi\}= \mathbb{S}_1 \times \mathbb{S}_1,$ known as  flat torus. 
Taking  partial derivatives of $X$ with respect to $\phi$, and $\theta$ are 
$$ X_{\phi}=\{ -(R+r\cos{\theta})\sin{\phi}, (R+r\cos{\theta})\cos{\phi}, 0 \}$$ and
$$X_{\theta}=\{  -r\sin{\theta} \cos{\phi}, -r\sin{\theta}\sin{\phi}, r\cos{\theta}  \}$$ respectively. Hence, the coefficients of the first fundamental form are
 \begin{equation}
 \begin{aligned}
     E=\langle X_{\phi},X_{\phi}\rangle &= (R+r\cos{\theta})^2\\
    F=\langle X_{\phi},X_{\theta}\rangle &= 0 \\
     G=\langle X_{\theta},X_{\theta}\rangle &= r^2
 \end{aligned}
 \label{fff cof}
 \end{equation}
leading to the area element of the curved torus (Equation-\ref{torus para equn}) 
\begin{equation}
    dA=r(R+r\cos{\theta})~d\phi~d\theta
    \label{torus area element}
\end{equation}
 Now using this area-element \cite{biswas2024changepoint} have introduced a new concept ``square of an angle'' denoted by $A_C^{(0)}(\theta)$, which is the minimum area between $(0,0)$ to  $(\theta, \theta)$ on the curved torus. We use this new  area-based measurement, $A_C^{(0)}(\theta)$ to construct the least-square equivalent condition for the linear-circular regression.\\

\subsection*{Motivating data example:}
Cryptocurrency is an innovative form of digital asset that operates as a decentralised medium of exchange over computer networks, without the need for any central authority like governments or banks. Cryptocurrencies utilize cryptographic techniques to ensure the security of transactions, regulate the generation of new units, and authenticate the transfer of assets by using blockchain technology. The decentralized nature of cryptocurrencies facilitates peer-to-peer transactions and fosters financial inclusion by granting unbanked populations access to financial services. Notable instances include Bitcoin, the groundbreaking digital money, and Ethereum, which introduced smart contracts that allow for programmable and automated transactions. The volatility of cryptocurrency markets and speculative character pose distinct difficulties and opportunities for financial modeling, risk assessment, and regulation altogether.

Extreme cryptocurrency prices whether highs or lows reflect shifts in supply-demand dynamics. Highs and lows throughout the day offer critical insights into market sentiment: a high price signals strong buying interest, whereas a low suggests selling pressure. Open and close prices are especially significant, as they represent the sentiment of the market at the beginning and end of the trading day. Open prices set the initial tone for traders, while close prices can drive after-market decisions and predict the next day’s momentum.
Market sentiment also plays a crucial role, with negative news leading to declines and positive endorsements or news triggering price surges. After significant price spikes, investors may sell to secure profits, often causing temporary price decreases.
Extreme price events create arbitrage opportunities, allowing traders to profit from price discrepancies across exchanges. The timing of these highs and lows also influences traders' psychology; early-day lows may prompt caution, while late-day lows encourage position reassessment. Conversely, early-day highs can foster optimism, while late-day highs may lead to profit-taking. Recognizing when these price extremes occur can help traders identify crucial support and resistance levels, enhancing strategy. Liquidity also impacts price stability; low liquidity periods can heighten price volatility, while high liquidity provides stability.
By understanding the timing and significance of high, low, open, and close prices, traders can refine their strategies. 

\subsection*{Challenges with this data:} Cryptocurrency markets operate continuously, 24 hours a day, 365 days a year, due to their decentralized architecture that enables trading across multiple global exchanges without fixed opening and closing times. Unlike traditional financial markets, which pause for holidays and specific hours, the cryptocurrency market allows traders to respond instantly to market events regardless of time zone or physical location. This round-the-clock environment introduces heightened volatility, as prices can shift rapidly with the unceasing flow of market data and trading activity. Consequently, sophisticated risk management, real-time data analytics, and algorithmic trading techniques are necessary to capitalize on the continuous fluctuations in market conditions.  The continuous availability of data introduces a cyclical pattern in the timestamps, making linear-circular regression a robust alternative to conventional linear methods that fail to address this cyclicity. In sum, linear-circular regression is a powerful tool for traders and analysts aiming to optimize strategies in a perpetually active market environment, allowing for enhanced decision-making based on the interplay of price and time.

This paper presents a novel regression model that connects angular responses with linear predictor using a generalized M\"{o}bius transformation, which distinctively maps the real axis to the unit circle. The model introduces an innovative area-based loss function on the surface of a curved torus, obtained using the intrinsic geometry of it. This area-based loss function becomes a natural extension of least square approach to estimate the parameters when the response variable is circular. This semi-parametric approach does not rely on specific distributional assumptions for the angular error, providing greater flexibility and broader applicability.

The structure of the paper is as follows: Section \ref{regression model sec} introduces the proposed regression model, formulated through a generalized M\"{o}biusmap that maps the real axis onto the circle, along with the non-parametric assumptions of angular error. In Section \ref{loss fun sec}, we first define a area-based loss function for parameter estimation using the intrinsic geometry of a curved torus. The section note only  provides the point-estimation method of the parameters  and but also gives  bootstrap procedures for calculating confidence and prediction intervals for a new data. Section \ref{simulation sec} presents extensive simulation studies. A comparative analysis with existing models is reported in Section \ref{comparison model sec}. Finally, before concluding in Section \ref{conclusion}, we analyze real-world datasets related to Bitcoin and Ethereum in Section \ref{data analysis sec}.

\section{ Regression model} \label{regression model sec} 

\subsection{Regression Curve:}
The M\"{o}bius transformations that map the upper half of the complex plane $\mathbb{C},$ where Im$(z) > 0$ onto the open disk $ \{w \in \mathbb{C}:|w| < 1\} $ and the boundary Im$(z) = 0$ of the half-plane onto the boundary of the disk with $|w| = 1.$  Such a M\"{o}bius transformation mathematically can be expressed as 
\begin{equation}
    w=M(z;\beta_0,\beta_1)=\beta_0~~ \frac{z-\beta_1}{z-\bar{\beta_1}},
    \label{full model}
\end{equation}
where $\beta_0 \in \mathbb{S}_1=\{w: |w|=1\} $, the unit circle, and $\beta_1 \in \mathbb{C}$. Here $\beta_0$ works as a rotation parameter. The point $\beta_1 \in \mathbb{C}$, which lies in the upper half-plane, is mapped (Equation-\ref{full model}) to the origin $w=0$ in the $w$-plane. This choice of $\beta_1$ 
  as a reference point affects the ``center" of the transformation in the 
$z$-plane, effectively defining where the origin in the $w$-plane corresponds to the $z$-plane \cite[see][pp. 324-325, Sec. 95]{brown2009complex}.

Since we are interested in linear-circular regression, that is the predictor would be a real variable and the response would be a circular variable, we must consider the map for the boundary of the upper-half plane, Im$(z) = 0$ which is $ \mathbb{R}$,  onto the boundary of the disk, $|w| = 1$ which is $ \mathbb{S}_1$. Hence, using Equation-\ref{full model} we can write 

\begin{equation}
    M(x;\beta_0,\beta_1)=\beta_0~~ \frac{x-\beta_1}{x-\bar{\beta_1}},
    \label{full real model}
\end{equation}
where $\beta_0 \in \mathbb{S}_1 $, and $\beta_1 \in \mathbb{C}$. 
This transformation can be decomposed into a sequence of four fundamental types of maps:
\begin{enumerate}
    \item \textbf{Translation:} Shift \(x\) by \(b\): $x \rightarrow x +b.$

    \item \textbf{Inversion:} Apply an inversion: $x \rightarrow  \frac{1}{x}.$

    \item \textbf{Scaling:} Scale by \(\alpha\): $x \rightarrow  \alpha x.$
\end{enumerate}
It is important to note that these transformations demonstrate the action of the group on the complex plane. Hence, the map in Equation-\ref{full real model} can be expressed as 
$$ M(x;\beta_0,\beta_1)=\beta_0~~ \left[ 1- \frac{\beta_1-\bar{\beta_1}}{x-\bar{\beta_1}} \right].$$
Here $\beta_0$ works as a rotation parameter as well. The interpretation $\beta_1$ is not straightforward, one important observation is if $|\beta_1|\rightarrow 0$ then $ \arg{[M(x;\beta_0,\beta_1)]}\rightarrow 0~\text{(radian)}.$
Now, using Equation-\ref{full real model} the regression model for the response variable $\theta \in [0, 2\pi)$ and the predictor $x \in \mathbb{R}$ can be defined as
\begin{eqnarray}
       \theta&=&  \arg{(M(x;\beta_0,\beta_1))} +\epsilon  \mod{ 2\pi}  \\ \nonumber
       &\equiv&  g(x;\beta_0,\beta_1)+\epsilon \mod 2\pi
       ,
       \label{the model}
\end{eqnarray}
where $g(x;\beta_0,\beta_1)$ is the link function with   $\beta_0 \in \mathbb{S}_1=\{w: |w|=1\} $, $\beta_1 \in \mathbb{C}$, and $\epsilon$ is a random angular error with zero mean belongs to $[0, 2\pi)$. If we put the estimated value of $\beta_0, \beta_1$, we would get the regression curve. 

\begin{figure*}[h!]
	\centering
\includegraphics[trim= 20 20 20 20, clip, width=0.9\textwidth,height=.75\textwidth]{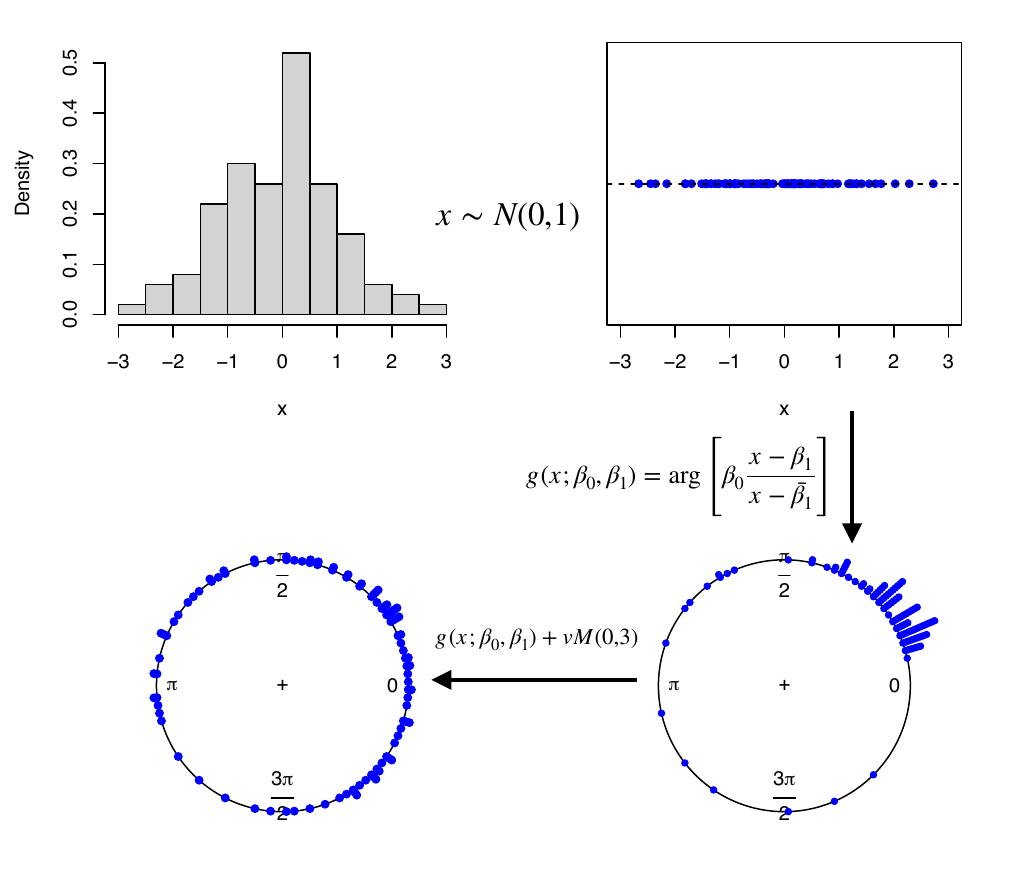}
	\caption{Top-left: Histogram of the predictors sampled from a standard normal distribution.  
Top-right: Plot of the predictor values on the real axis.  Bottom-right: Plot of the expected responses through the M\"{o}biusmap with $\beta_0=1,\beta_1=1.5+0.5i$ on the unit circle.  Bottom-left: Plot of the responses with angular error, $\epsilon$ drawn from a von Mises distribution with zero mean direction and concentration, $\kappa=3$ or $vM(0,3)$.}
    \label{fig: linear_cp_plot}
\end{figure*}

It is worth noting that the distribution of the angular error is free from any model assumption apart from its zero mean direction with fixed variance over samples. Here, we need to estimate only the parameters from the M\"{o}biusmap. Hence it is a \textit{semi-parametric linear-circular} regression model.\\

\section{Loss Function and  Estimation} \label{loss fun sec}
In least-squares regression analysis, the mean squared error (MSE) is employed as the loss function to quantify the average squared difference between the predicted values and the actual observed values. By minimizing the MSE, we estimate the model parameters that provides best fit to the data. This minimization ensures that the predicted values are as close as possible to the observed ones, reducing the overall error in the regression model. Now, for observed values \( y_i \) and predicted values \( \hat{y}_i \) (\( i = 1, 2, \cdots, n \)), the MSE can be computed as:

\[
\text{MSE} = \frac{1}{n} \sum_{i=1}^{n} \left(y_i - \hat{y}_i\right)^2= \displaystyle \frac{1}{n} \sum_{i=1}^{n} |\mathbb{Y}_{i}|^{2},
\]
where each \( |\mathbb{Y}_{i}|^2 = |\left(y_i - \hat{y}_i\right)|^2 \), for \( i = 1, 2, \cdots, n \)  geometrically represents the mean of random squared areas with arm length $|\mathbb{Y}_{i}|$ in \( \mathbb{R}^2 \) for (\( i = 1, 2, \cdots, n \)) .

\subsection{Loss function:} 
Let \((\theta_i, x_i) \in (\mathbb{S}_1 \times \mathbb{R})\) for \(i=1,2,\cdots,n\) represent \(n\) observed data points, where \(\mathbb{S}_1\) denotes the unit circle and \(\mathbb{R}\) represents real-valued predictors. The predicted values for \(\theta_i\) are given by
\[
\hat{\theta}_i = g(x_i; \hat{ \beta_0}, \hat{ \beta_1}) \mod 2\pi
\]
for an estimate $(\hat{ \beta_0}, \hat{ \beta_1})$ of the parameters.
The residuals, representing the angular difference between observed and predicted values, are defined as:  
\[
\psi_i = (\theta_i - \hat{\theta}_i) \mod 2\pi,
\]
then motivated by this MSE and its geometric interpretation discussed above \cite{biswas2024changepoint} have drawn the analogy to define  the ``square of an angle''.  We use the same to define the  \textit{mean square  angle error} \textit{(MSAE)} for the proposed regression model. The estimated values of the parameters $(\beta_0, \beta_1)$ are obtained by minimizing the MSAE. Hence, the loss function can be  written as 

\begin{eqnarray}
    \mathcal{L}(\beta_0, \beta_1)= \min_{\beta_0,\beta_1} \frac{1}{n}\sum_{i=1}^{n} A_{0}^{C}(\psi_i).
    \label{loss function}
\end{eqnarray}

Since the loss function does not have a closed-form solution for the parameters, some numerical optimization techniques can be employed to minimize the loss function presented in Equation-\(\ref{loss function}\) and obtain the estimated values of \(\beta_0\) and \(\beta_1\). In this analysis, we utilized the ``L-BFGS-B" numerical optimization method, which is well-suited for problems with bounds. This approach ensures precise minimization of the loss function, enabling efficient parameter estimation for the proposed model.\\

\subsection{Confidence and Prediction interval:}
Let $(\theta_i, x_i) \in (\mathbb{S}_1 \times \mathbb{R})$ for $i=1,2, \cdots,n$ be $n$ observed data points. The regression model is given by 
\begin{equation}
    \theta_i=g(x_i;\beta_0,\beta_1) +\epsilon_i \mod 2\pi.
    \label{model_equation}
\end{equation}

where $g(x_i;\beta_0,\beta_1)$ is the link function with   $\beta_0 \in \mathbb{S}_1 $, $\beta_1 \in \mathbb{C}$, and $\epsilon_i$ is a random angular error with zero mean and fixed variance for $i=1, \cdots,n.$ Now we provide the method to obtain $95\%$ confidence interval (CI) for $E(\theta_i|x_i)$ for $i=1, \cdots,n.$

\rule{\textwidth}{1pt}  

\begin{center}
Bootstrap algorithm for calculating CI.
\end{center}

\rule{\textwidth}{0.5pt}  

\begin{center}
\begin{enumerate}
    \item True mean: $\mu_{\theta_i|x_i}=E(\theta_i|x_i)=g(x_i;\beta_0,\beta_1) ~~\text{for~~} i=1, \cdots,n.$
    \label{true_mean}\\

    \item We can write  
   $ \mu_{\theta_j|x_j}=E(\theta_j|x_j)=g(x_j;\beta_0,\beta_1)$ using (\ref{true_mean}) \\
   where $x_j \in \mathbb{R} $ be any predictor variable. \\

   \item The predicted values are $\hat{\theta_i}=g(x_i;\hat{\beta}_0,\hat{\beta}_1)~~\text{for~~} i=1, \cdots,n$ where $\hat{\beta}_0,\hat{\beta}_1$ are the estimated values of the parameters $\beta_0,\beta_1$.\\

   \item The predicted value at $x_j$ is $\hat{\mu}_{\theta_j|x_j}=E(\hat{\theta}_j|x_j)=g(x_j;\hat{\beta}_0,\hat{\beta}_1)
    \label{estimated_mean_new_point}$\\

    \item Get angular error: $e_i=(\theta_i-\hat{\theta_i}) \mod 2\pi ~~\text{for~~} i=1, \cdots,n.$\\

    \item Let $B$ be the total number of bootstraps and $\textbf{e}=(e_1,e_2,\cdots,e_n).$
    \label{error}
    \item For $b=1,2,\ldots,B:$
     \begin{enumerate}
    \renewcommand{\labelenumi}{\roman{enumi}}
    \item $y^{(b)}_i=[g(x_i; \hat{\beta}_0,\hat{\beta}_1 )$ + a sample with replacement from $\textbf{e}] \mod 2\pi ~~\text{for~~} i=1, \cdots,n.$
    \item Obtained the $b$-th bootstrap estimated values of $\beta_0,\beta_1$, and call it $\hat{\hat{\beta}}^{(b)}_{0},\hat{\hat{\beta}}^{(b)}_{1}$.
    \item Compute the response for the predictor $x_j$ from $(2)$ as $\hat{\theta}_{j}^{b}=g(x_j;\hat{\hat{\beta}}^{(b)}_{0},\hat{\hat{\beta}}^{(b)}_{1})$
\end{enumerate}

\item Calculate the $95\%$ CI from the bootstrap responses: $(\hat{\theta}_{j}^{1},\hat{\theta}_{j}^{2},\cdots, \hat{\theta}_{j}^{B}).$
\end{enumerate}

\end{center}

\rule{\textwidth}{1pt}  

Check whether $\mu_{\theta_j|x_j}=E(\theta_j|x_j)$ from (2) and $\hat{\mu}_{\theta_j|x_j}=E(\hat{\theta}_j|x_j)$ and (4) fall inside the confidence interval or not for a large number of iterations to compute the coverage probability.\\

Now, we provide the bootstrap algorithm for prediction interval.

\rule{\textwidth}{1pt}  

\begin{center}
Algorithm for calculating prediction interval
\end{center}

\rule{\textwidth}{0.5pt}  

\begin{center}
\begin{enumerate}
    \item Let $x_j \in \mathbb{R} $ be any new predictor variable and corresponding response variable\\ $ \theta_j=g(x_j;\beta_0,\beta_1)+ \epsilon_j~ \mod 2\pi$ using Equation-\ref{model_equation}.
   \\

   \item The predicted values are $\hat{\theta_i}=g(x_i;\hat{\beta}_0,\hat{\beta}_1)~~\text{for~~} i=1, \cdots,n$ where $\hat{\beta}_0,\hat{\beta}_1$ are the estimated values of the parameters $\beta_0,\beta_1$.\\

   \item The predicted value for $x_j$ is $\hat{\theta}_j=g(x_j;\hat{\beta}_0,\hat{\beta}_1)
    $\\

    \item Get angular error: $e_i=(\theta_i-\hat{\theta_i}) \mod 2\pi ~~\text{for~~} i=1, \cdots,n.$\\

    \item Let $B$ be the total number of bootstraps and $\textbf{e}=(e_1,e_2,\cdots,e_n).$
    \item For $b=1,2,\ldots,B:$
     \begin{enumerate}
    \renewcommand{\labelenumi}{\roman{enumi}}
    \item $y^{(b)}_i=[g(x_i; \hat{\beta}_0,\hat{\beta}_1 )$ + a sample with replacement from $\textbf{e}] \mod 2\pi ~~\text{for~~} i=1, \cdots,n.$
    \item Obtained the $b$-th bootstrap estimated values of $\beta_0,\beta_1$, and call it $\hat{\hat{\beta}}^{(b)}_{0},\hat{\hat{\beta}}^{(b)}_{1}$.
    \item Compute the response for the predictor $x_j$ from $(2)$ as \\ $\hat{\theta}_{j}^{b}=g(x_j;\hat{\hat{\beta}}^{(b)}_{0},\hat{\hat{\beta}}^{(b)}_{1})$+ a sample with replacement from $\textbf{e}] \mod 2\pi.$
\end{enumerate}

\item Calculate the $95\%$ CI from the bootstrap responses: $(\hat{\theta}_{j}^{1},\hat{\theta}_{j}^{2},\cdots, \hat{\theta}_{j}^{B}).$ \\
\end{enumerate}

\end{center}

\rule{\textwidth}{1pt}  

 Check whether $ \theta_j=g(x_j;\beta_0,\beta_1)+ \epsilon_j~ \mod 2\pi$ from (1) and $\hat{\theta}_j=g(x_j;\hat{\beta}_0,\hat{\beta}_1)$ and (3) fall inside the interval or not for a large number of iterations to validate the probability associated to the prediction interval.\\

\begin{figure}[h!]
    \centering
    \subfloat[]{%
        {\includegraphics[trim= 80 90 90 60, clip,width=0.45\textwidth, height=0.3\textwidth]{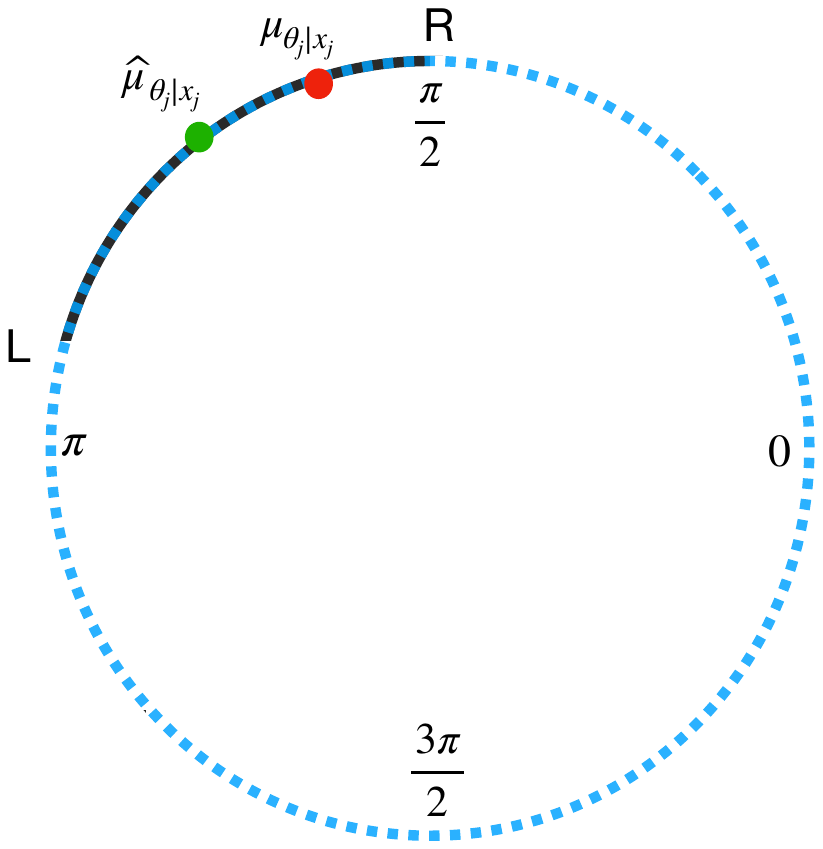}}}\hspace{5pt}
    \subfloat[ ]{%
        {\includegraphics[trim= 80 90 90 60, clip,width=0.45\textwidth, height=0.3\textwidth]{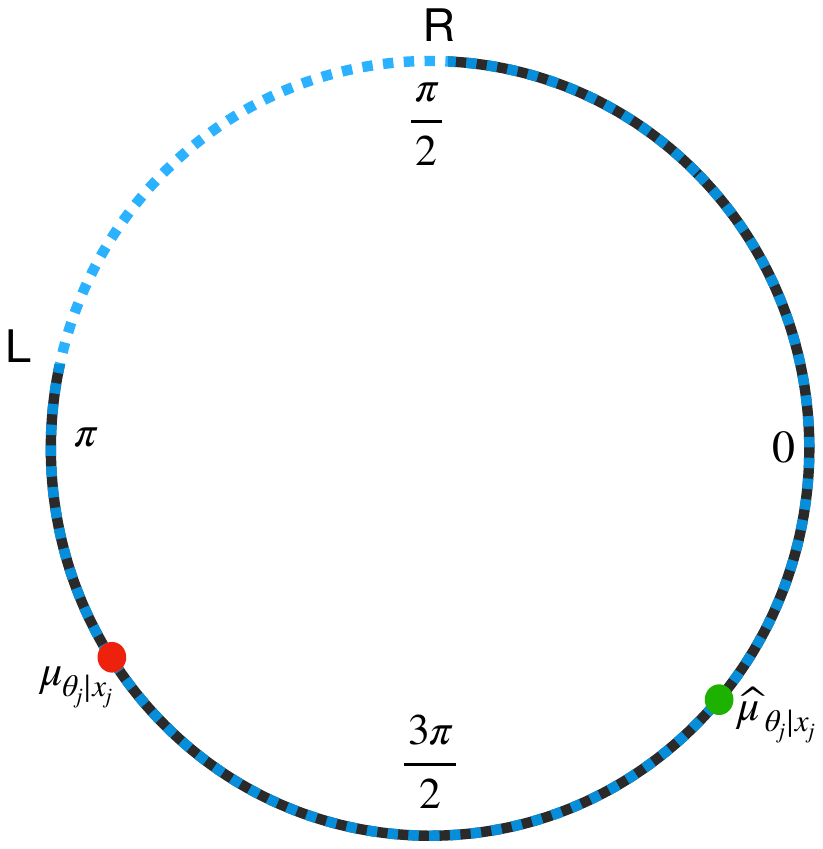}}}
    \caption{ The black arc on the circle represents the CI which englobes the true and the predicted conditional means.  The possible depictions of the confidence intervals on the circle can be seen in (A) and  (B).}
     \label{confidence interval}
\end{figure}

\textbf{Note:} To calculate this CI for both the algorithms, we have used \textit{quantile.circular} function in the package \textit{circular} by \cite{lund2017package} which is a widely used library in \textit{R} programming language for circular data analysis. The pictorial representation of the same is given in Figure-\ref{confidence interval}.

\section{ Simulation} \label{simulation sec}
In this section, we conduct a detailed simulation study to evaluate the performance of the proposed linear-circular regression model. The study examines the model under varying parameter specifications across three different sample sizes \((n)\) to assess its robustness and effectiveness. Specifically, we consider \(n = 50\) to explore small-sample behavior, \(n = 150\) for moderate sample size effects, and \(n = 500\) to study large-sample performance.

\begin{table}[t]
\centering
\resizebox{1\linewidth}{!}{%
  \begin{tabular}{|c|c|c|c|c|c|l|}
\hline
& & \multicolumn{3}{ c| }{Parameters} \\ \hline
Sample size& Concentration parameter& $b_0=0$ & $b_1=1.5$ & $b_2=0.5$  \\ \hline
\multicolumn{1}{ |c  }{\multirow{2}{*}{$n=50$} } &
\multicolumn{1}{ |c| }{$\kappa=0.5$} & 0.04382 (0.3530) & 1.2674 (0.0055)  & 0.4333 (0.0058)    \\ 
\multicolumn{1}{ |c  }{}                        &
\multicolumn{1}{ |c| }{$\kappa=1$}  & 0.0058 (0.1346) & 1.3740 ( 0.0040) & 0.4914 (0.0042)   \\ 
\multicolumn{1}{ |c  }{}                        &
\multicolumn{1}{ |c| }{$\kappa=10$}  & 0.0058 (0.0524) & 1.4880 (0.0009) & 0.4900 (0.0008)    \\ 
\hline
\multicolumn{1}{ |c  }{\multirow{2}{*}{$n=100$} } &
\multicolumn{1}{ |c| }{$\kappa=0.5$} & 0.0015 (0.2800) & 1.3945 (0.0063)  & 0.4592 (0.0067)   \\ 
\multicolumn{1}{ |c  }{}                        &
\multicolumn{1}{ |c| }{$\kappa=1$}  & 0.0117 (0.1346) & 1.4717 ( 0.0052)& 0.4758 (0.0011)    \\ 
\multicolumn{1}{ |c  }{}                        &
\multicolumn{1}{ |c| }{$\kappa=10$}  & -0.0002 (0.0148) & 1.5001 (0.0004)  & 0.50003 (0.0004)   \\ 
\hline
\multicolumn{1}{ |c  }{\multirow{2}{*}{$n=500$} } &
\multicolumn{1}{ |c| }{$\kappa=0.5$} & 0.0096 (0.1300) & 1.4328 (0.0031)  & 0.4717 (0.0027)   \\ 
\multicolumn{1}{ |c  }{}                        &
\multicolumn{1}{ |c| }{$\kappa=1$}  & -0.0002 (0.0692) & 1.4976 ( 0.0011)& 0.4999 (0.0011)    \\ 
\multicolumn{1}{ |c  }{}                        &
\multicolumn{1}{ |c| }{$\kappa=10$}  & -0.0001 (0.0307) & 1.499672 (0.0002)  & 0.4999 (0.0002)   \\ 
\hline
\end{tabular}}
\vspace{0.3cm}
\caption{The parameter estimates of \((b_0, b_1, b_2) = (0, 1.5, 0.5)\), with standard errors indicated in parentheses. These estimates were obtained when the predictors, sampled with varying sizes, followed a standard normal distribution, while the angular errors were drawn from a von Mises distribution with a zero mean direction and varying concentration parameters.}
\label{table:von_estimated_parameters_table_0_1.5_0.5}
\end{table}

\begin{table}[h!]
\centering
\resizebox{1\linewidth}{!}{%
  \begin{tabular}{|c|c|c|c|c|c|l|}
\hline
& & \multicolumn{3}{ c| }{Parameters} \\ \hline
Sample size& Concentration parameter& $b_0=\pi/6=0.52359$ & $b_1=-0.7$ & $b_2=2.4$  \\ \hline
\multicolumn{1}{ |c  }{\multirow{2}{*}{$n=50$} } &
\multicolumn{1}{ |c| }{$\kappa=0.5$} & 0.6189 (0.5837)  & -0.5468 (0.0198) & 2.2522 (0.0166)   \\ 
\multicolumn{1}{ |c  }{}                        &
\multicolumn{1}{ |c| }{$\kappa=1$}  & 0.5616 (0.3770) & -0.6344 (0.0097) & 2.3540 (0.0093)    \\ 
\multicolumn{1}{ |c  }{}                        &
\multicolumn{1}{ |c| }{$\kappa=10$}  & 0.5242 (0.0762) & -0.6982 (0.0016) & 2.4016 (0.0019)     \\ 
\hline

\multicolumn{1}{ |c  }{\multirow{2}{*}{$n=100$} } &
\multicolumn{1}{ |c| }{$\kappa=0.5$} & 0.5567(0.5427)  & -0.5587 (0.0170) & 2.3522 (0.0155)   \\ 
\multicolumn{1}{ |c  }{}                        &
\multicolumn{1}{ |c| }{$\kappa=1$}  & 0.5260 (0.2702) & -0.6939 (0.0068) & 2.3916 (0.0062)    \\ 
\multicolumn{1}{ |c  }{}                        &
\multicolumn{1}{ |c| }{$\kappa=10$}  & 0.5242 (0.0762) & -0.6982 (0.0016) & 2.4016 (0.0019)     \\  
\hline

\multicolumn{1}{ |c  }{\multirow{2}{*}{$n=500$} } &
\multicolumn{1}{ |c| }{$\kappa=0.5$} & 0.5445 (0.2093) & -0.6691 (0.0007) & 2.4020 (0.0007)    \\ 
\multicolumn{1}{ |c  }{}                        &
\multicolumn{1}{ |c| }{$\kappa=1$} & 0.5315 (0.1276) & -0.6906 (0.0004) & 2.4009 (0.0003)    \\ 
\multicolumn{1}{ |c  }{}                        &
\multicolumn{1}{ |c| }{$\kappa=10$}  & 0.5237 (0.02325) & -0.6998 (0.0007) & 2.4002 (0.0007)    \\ 
\hline
\end{tabular}}
\vspace{0.3cm}
\caption{The parameter estimates of \((b_0, b_1, b_2) = (\pi/6, -0.7, 2.4)\), with standard errors indicated in parentheses. These estimates were derived when the predictors, sampled with varying sizes, followed a standard normal distribution, and the angular errors were drawn from a von Mises distribution with a zero mean direction and varying concentration parameters.}
\label{table:von_estimated_parameters_table_pi/6_-0.7_2.4}
\end{table}

\begin{table}[t]
\centering
\resizebox{1\linewidth}{!}{%
  \begin{tabular}{|c|c|c|c|c|c|l|}
\hline
& & \multicolumn{3}{ c| }{Parameters} \\ \hline
Sample size& Concentration parameter& $b_0=0$ & $b_1=1.5$ & $b_2=0.5$  \\ \hline
\multicolumn{1}{ |c  }{\multirow{2}{*}{$n=50$} } &
\multicolumn{1}{ |c| }{$\rho=0.3$} & 0.0743 (0.3978) & 1.3164 (0.0045) & 0.4066 (0.0044)   \\ 
\multicolumn{1}{ |c  }{}                        &
\multicolumn{1}{ |c| }{$\rho=0.6$}  & 0.0033 (0.1816) & 1.4552(0.0021) & 0.4938 (0.0021)     \\ 
\multicolumn{1}{ |c  }{}                        &
\multicolumn{1}{ |c| }{$\rho=0.8$}  & 0.0048 (0.0914) & 1.4732 (0.0017) & 0.4913 (0.0018)   \\ 
\hline
\multicolumn{1}{ |c  }{\multirow{2}{*}{$n=100$} } &
\multicolumn{1}{ |c| }{$\rho=0.3$} & -0.0114 (0.2351) & 1.3835 (0.0052) & 0.4578 (0.0020)   \\ 
\multicolumn{1}{ |c  }{}                        &
\multicolumn{1}{ |c| }{$\rho=0.6$}  & 0.0182 (0.1210) & 1.4696(0.0029) & 0.4679 (0.0046)     \\ 
\multicolumn{1}{ |c  }{}                        &
\multicolumn{1}{ |c| }{$\rho=0.8$}  & -0.0015 (0.0472) & 1.4990 (0.0007) & 0.5008 (0.0008)    \\ 
\hline
\multicolumn{1}{ |c  }{\multirow{2}{*}{$n=500$} } &
\multicolumn{1}{ |c| }{$\rho=0.3$} & 0.0080 (0.1007) & 1.4809 (0.0022) & 0.4919 (0.0020)   \\ 
\multicolumn{1}{ |c  }{}                        &
\multicolumn{1}{ |c| }{$\rho=0.6$}  & 0.0001 (0.0431) & 1.5001 (0.0006) & 0.5001 (0.0007)     \\ 
\multicolumn{1}{ |c  }{}                        &
\multicolumn{1}{ |c| }{$\rho=0.8$}  & -0.0002 (0.0231) & 1.4999 (0.0003) & 0.5000 (0.0003)    \\ 
\hline
\end{tabular}}
\vspace{0.3cm}
\caption{The parameter estimates of \((b_0, b_1, b_2) = (0, 1.7, 0.5)\), with standard errors indicated in parentheses. These estimates were obtained when the predictors, sampled with varying sizes, followed a standard Cauchy distribution, and the angular errors were drawn from a wrapped Cauchy distribution with a zero mean direction and varying concentration parameters.}
\label{table: wc_estimated_parameters_table_0_1.5_0.5}
\end{table}

\begin{table}[t]
\centering
\resizebox{1\linewidth}{!}{%
  \begin{tabular}{|c|c|c|c|c|c|l|}
\hline
& & \multicolumn{3}{ c| }{Parameters} \\ \hline
Sample size& Concentration parameter& $b_0=\pi/6=0.52359$ & $b_1=-0.7$ & $b_2=2.4$  \\ \hline
\multicolumn{1}{ |c  }{\multirow{2}{*}{$n=50$} } &
\multicolumn{1}{ |c| }{$\rho=0.3$} & 0.5806 (0.5091) & -0.5718 (0.0016) & 2.1248 (0.2372)   \\ 
\multicolumn{1}{ |c  }{}                        &
\multicolumn{1}{ |c| }{$\rho=0.6$}  & 0.5347 (0.1945) & -0.6711 (0.0063) & 2.4250 (0.0076)     \\ 
\multicolumn{1}{ |c  }{}                        &
\multicolumn{1}{ |c| }{$\rho=0.8$}  & 0.5248 (0.1837) & -0.6886 (0.0047) & 2.3940 (0.0036)   \\ 
\hline
\multicolumn{1}{ |c  }{\multirow{2}{*}{$n=100$} } &
\multicolumn{1}{ |c| }{$\rho=0.3$} & 0.5690 (0.3510) & -0.6338 (0.0125) & 2.3527 (0.0020)   \\ 
\multicolumn{1}{ |c  }{}                        &
\multicolumn{1}{ |c| }{$\rho=0.6$}  & 0.5266 (0.1439) & -0.6894 (0.0036) & 2.3955 (0.0038)     \\ 
\multicolumn{1}{ |c  }{}                        &
\multicolumn{1}{ |c| }{$\rho=0.8$}  & 0.5263 (0.0863) & -0.6941 (0.0019) & 2.4025 (0.0022)    \\ 
\hline
\multicolumn{1}{ |c  }{\multirow{2}{*}{$n=500$} } &
\multicolumn{1}{ |c| }{$\rho=0.3$} & 0.5334 (0.1877) & -0.6826 (0.0060) & 2.4053 (0.0062)   \\ 
\multicolumn{1}{ |c  }{}                        &
\multicolumn{1}{ |c| }{$\rho=0.6$}  & 0.5256 (0.0760) & -0.6991 (0.0024) & 2.4008 (0.0024)     \\ 
\multicolumn{1}{ |c  }{}                        &
\multicolumn{1}{ |c| }{$\rho=0.8$}  & 0.5244 (0.0458) & -0.6992 (0.0001) & 2.3988 (0.0009)    \\ 
\hline
\end{tabular}}
\vspace{0.3cm}
\caption{The parameter estimates of \((b_0, b_1, b_2) = (\pi/6, -0.7, 2.4)\), with standard errors indicated in parentheses. These estimates were obtained when the predictors, sampled with varying sizes, followed a standard Cauchy distribution, and the angular errors were drawn from a wrapped Cauchy distribution with a zero mean direction and varying concentration parameters.}
\label{table: wc_estimated_parameters_table_pi/6_-0.7_2.4}
\end{table}

\begin{figure}[b]
    \centering
    \subfloat[]{%
        {\includegraphics[trim= 2 2 2 2, clip,width=0.4\textwidth, height=0.4\textwidth]{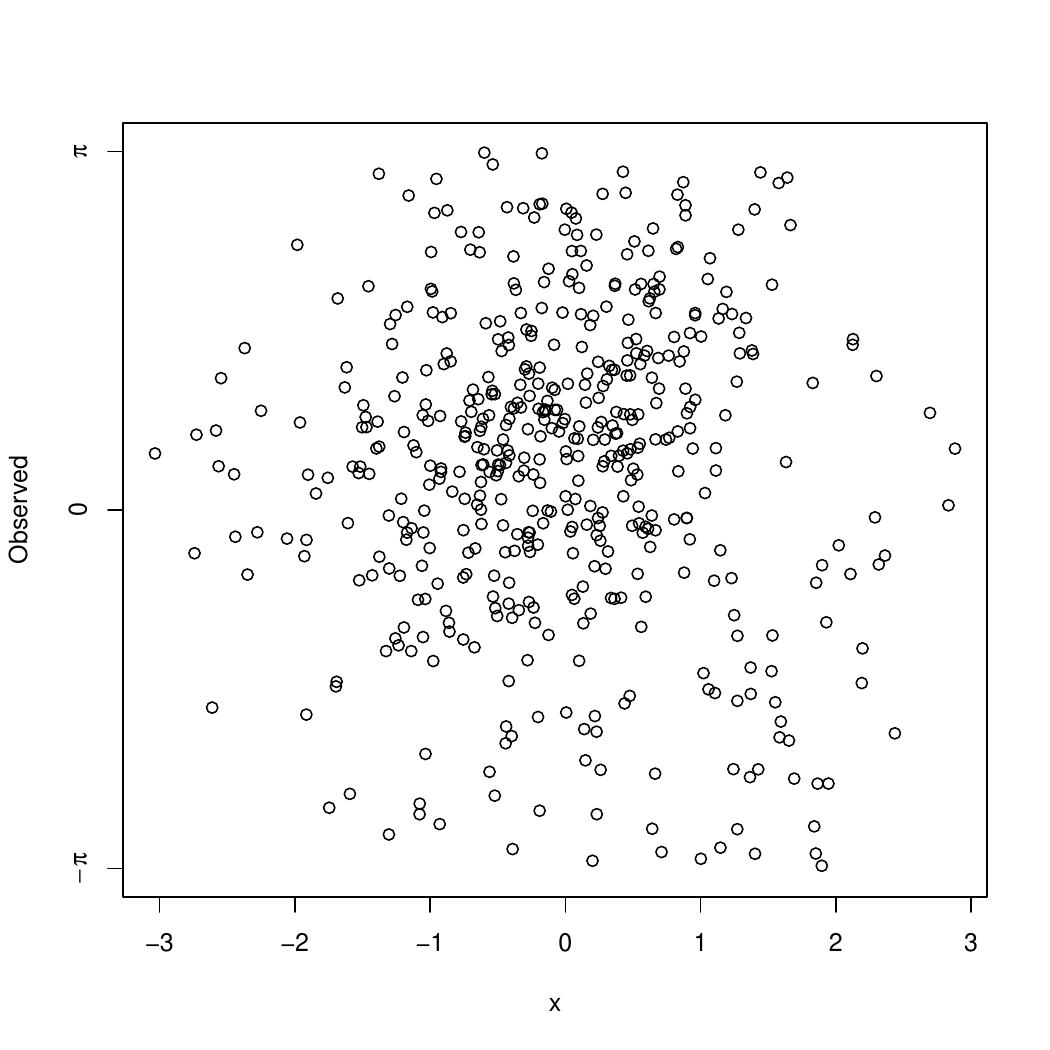}}}\hspace{5pt}
    \subfloat[ ]{%
        {\includegraphics[trim= 2 2 2 2, clip,width=0.4\textwidth, height=0.4\textwidth]{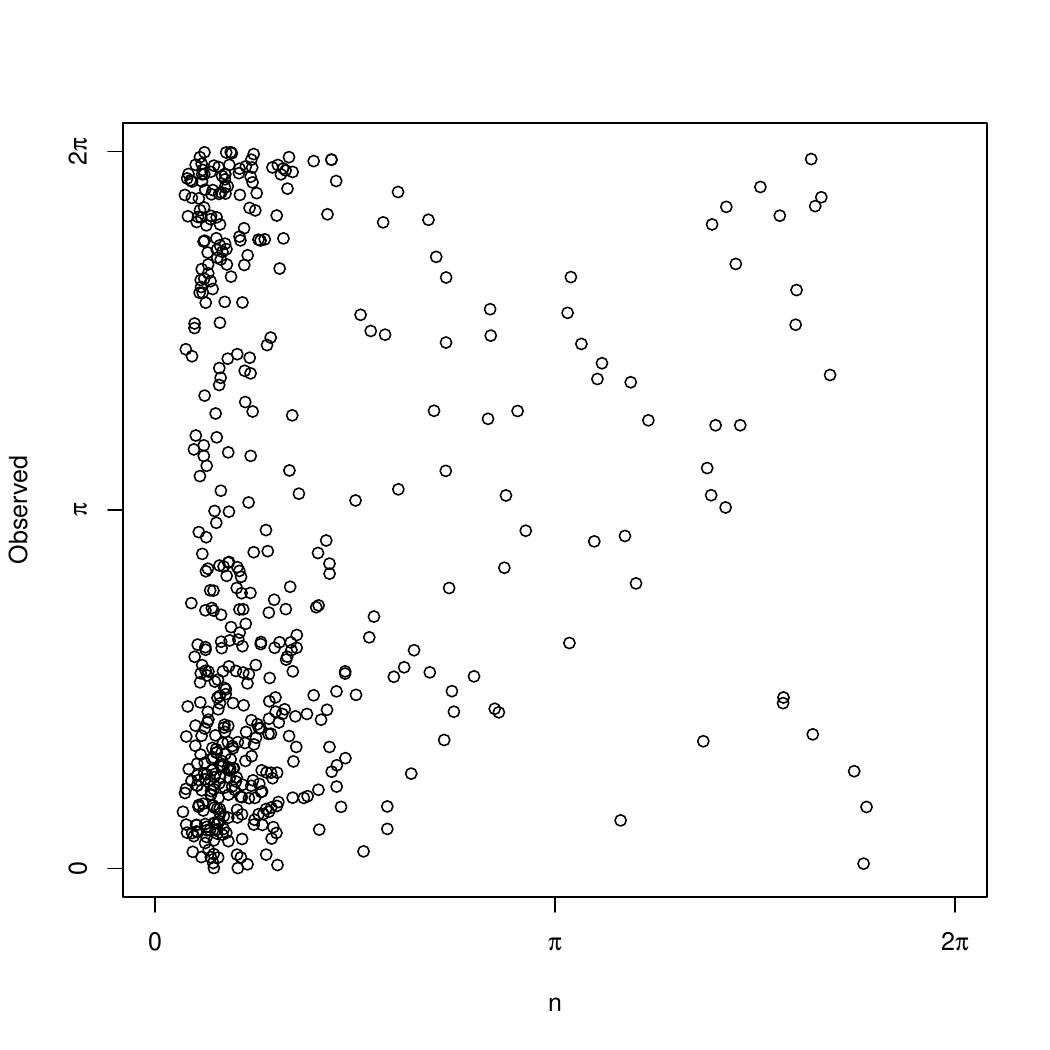}}}
   
    \subfloat[ ]{%
        {\includegraphics[trim= 2 2 2 2, clip,width=0.4\textwidth, height=0.4\textwidth]{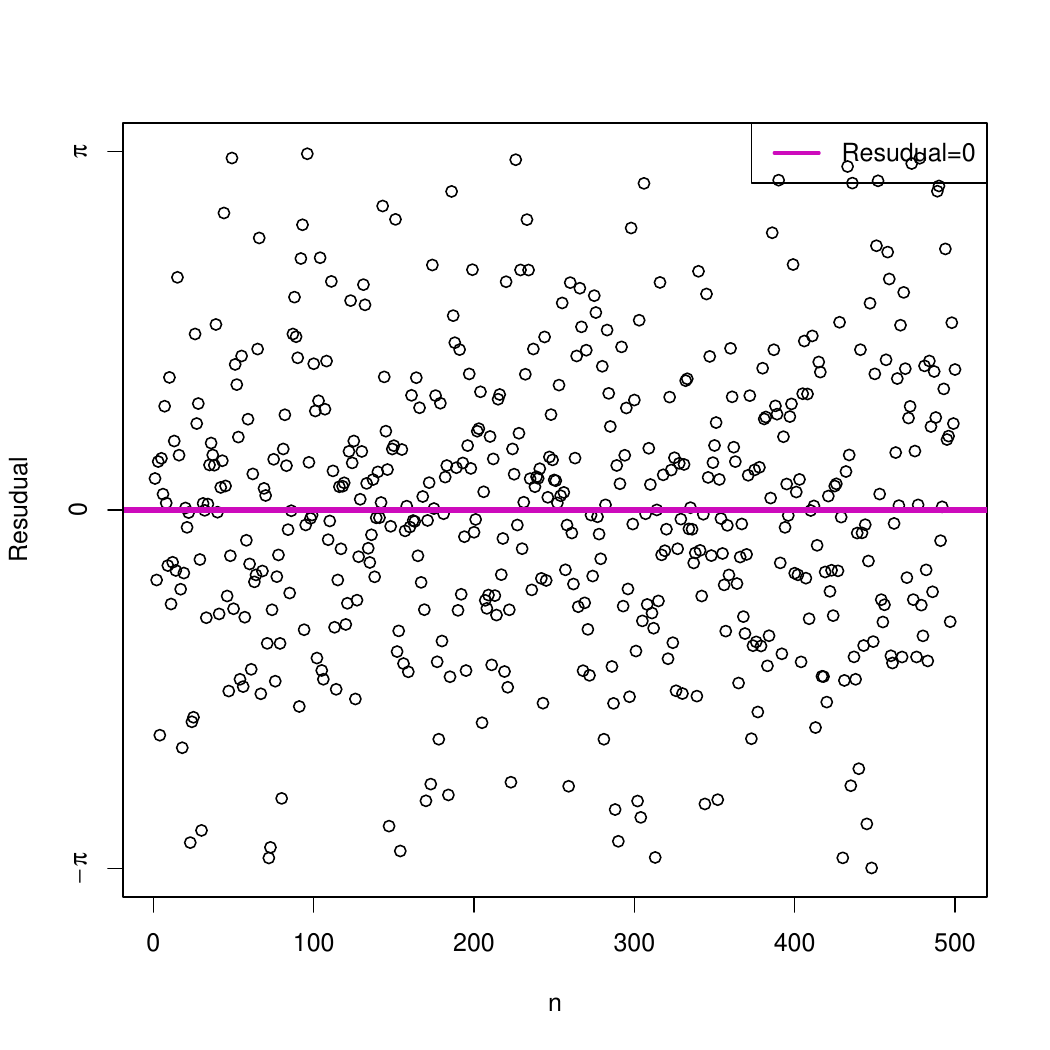}}}\vspace{5pt}
    \subfloat[ ]{%
        {\includegraphics[trim= 2 2 2 2, clip,width=0.4\textwidth, height=0.4\textwidth]{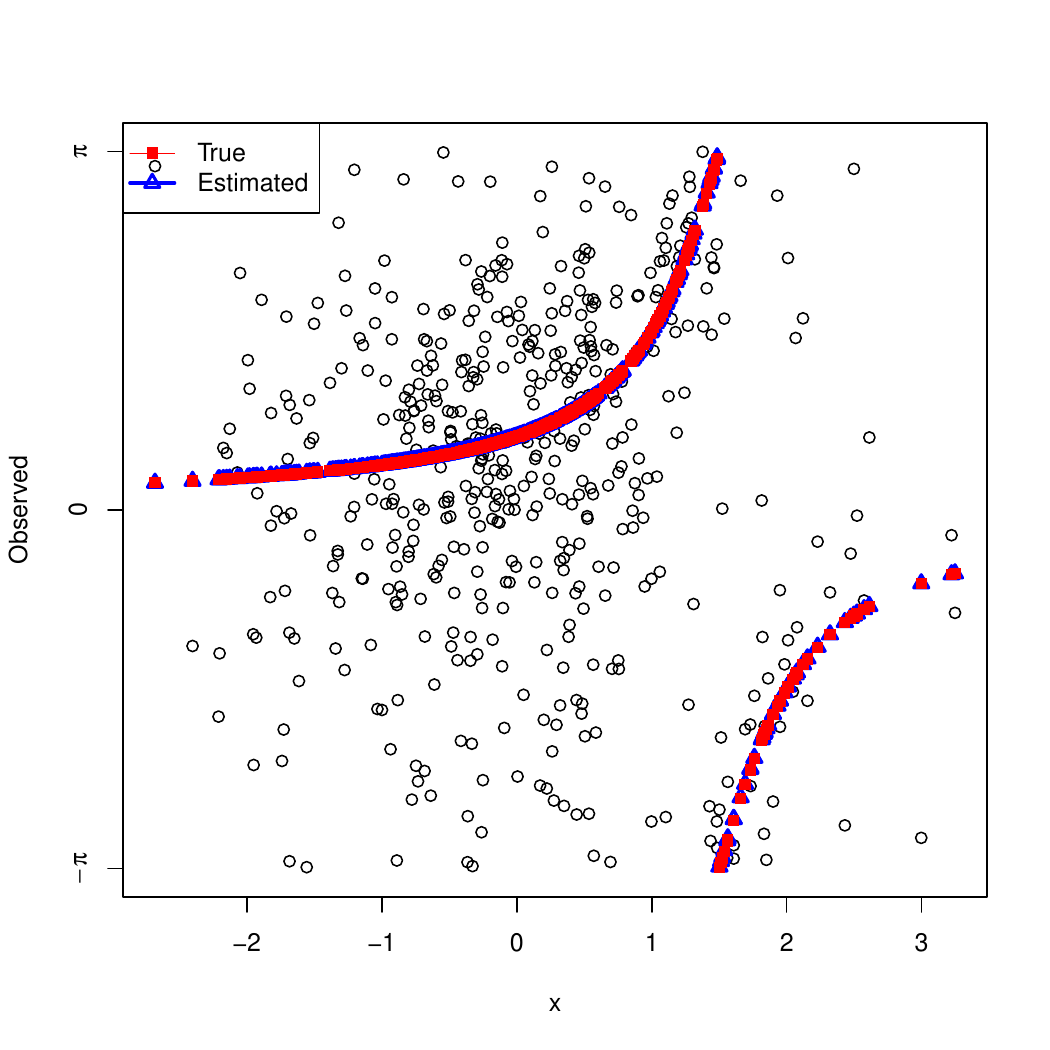}}}\vspace{5pt}
    \caption{(A) Scatter plot of the true simulated data, including angular error.  
(B) Scatter plot comparing the true data with the predicted data.  
(C) Plot of the residuals (restricted to the range \([-\pi, \pi]\) for enhanced visual clarity) with a reference line at Residual = 0.  
(D) Plot of the exact regression curve (red line with square) and the fitted regression curve (blue line with triangle) overlaid on the simulated dataset.}
     \label{simulated_plot_von_mises}
\end{figure}

To demonstrate that the angular error in the proposed regression model is not dependent on the choice of distribution, we incorporate two distinct angular error distributions: the von Mises distribution and the wrapped Cauchy distribution. For both distributions, the mean direction is set to \(\mu = 0\) (in radians), and simulations are performed under varying concentration parameters. This comprehensive setup allows us to investigate the behavior of the model under different distributional assumptions and sample size scenarios.

Since there exists a connection between the Normal and Cauchy distributions with the von Mises and wrapped Cauchy distributions, respectively, we generated the predictor \(x\) from the standard Normal distribution for simulations involving von Mises angular errors and from the standard Cauchy distribution for simulations involving wrapped Cauchy angular errors. However, it is important to note that the choice of distributions for both the predictor and the angular error is flexible, and other distributions can also be used depending on the application. For further details on these circular distributions and their properties, refer to \cite{mardia2000directional}.

The simulation results for the angular error models are summarized across different scenarios. Here, the parameter \(\beta_0\) is represented as \(\exp(i b_0)\), while \(\beta_1\) is expressed as \(b_1 + i b_2\).
When the angular error is drawn from a von Mises distribution, then the results are summarized in Table-\ref{table:von_estimated_parameters_table_0_1.5_0.5} for parameter values \(b_0=0\), \(b_1=1.5\), and \(b_2=0.5\), and in Table-\ref{table:von_estimated_parameters_table_pi/6_-0.7_2.4} for \(b_0=\frac{\pi}{6}\), \(b_1=-0.7\), and \(b_2=2.4\). Similarly, Table-\ref{table: wc_estimated_parameters_table_0_1.5_0.5} and Table-\ref{table: wc_estimated_parameters_table_pi/6_-0.7_2.4} present the simulation results when the angular error is from a wrapped Cauchy distribution. Additional simulations were performed for specific conditions. Under the von Mises distribution with \(b_0 = 0\), \(b_1 = -1.1\), \(b_2 = -1.8\), a sample size of \(n = 500\), \(\kappa = 5\), and zero mean direction, the parameter estimates were \(\hat{b}_0 = -0.0001\) (\(0.0301\)), \(\hat{b}_1 = -1.0997\) (\(0.0008\)), and \(\hat{b}_2 = -1.7996\) (\(0.0008\)). For the wrapped Cauchy distribution with \(b_0 = \frac{2\pi}{3}\) (\( \approx 2.0944\)), \(b_1 = -0.6\), \(b_2 = -0.8\), a sample size of \(n = 500\), \(\rho = 0.5\), and zero mean direction, the parameter estimates were \(\hat{b}_0 = 2.0938\) (\(0.0671\)), \(\hat{b}_1 = -0.5993\) (\(0.0009\)), and \(\hat{b}_2 = -0.8001\) (\(0.0009\)).
 For the parameter estimates, the mean values are computed over 10,000 simulations, with the standard errors reported in parentheses for \(b_1\) and \(b_2\). For \(b_0\), the circular mean of the estimates is presented, along with the circular variance shown in parentheses. This comprehensive representation highlights the behavior of the proposed model under various sample sizes and angular error distributions.

Figure-\ref{simulated_plot_von_mises}(A) displays the plot of the true simulated data. The x-axis represents predictors, \(x\), drawn from a standard normal distribution, while the y-axis represents the observed angles. These angles are derived as arguments of the transformed \(x\) through a M\"{o}biusmap with parameters \(b_0 = 0\), \(b_1 = 1.5\), and \(b_2 = 0.5\), combined with random angular errors sampled from a von Mises distribution with a zero mean direction and a concentration parameter \(\kappa = 1\). Figure-\ref{simulated_plot_von_mises}(B) shows the scatter plot of true data versus predicted data.
Figure-\ref{simulated_plot_von_mises}(C) illustrates the residual plot, showing the angular errors from the proposed linear-circular regression model applied to this simulated dataset. The residuals exhibit no  systematic pattern, indicating that the model effectively captures the relationship between the circular and linear components. Most residuals are concentrated near zero, consistent with the assumption of minimal angular error in a well-fitted model. Additionally, the residual spread is symmetric about the zero line. Similar to how residuals in linear regression follow a normal distribution under well-met assumptions, the residuals in this model follow a von Mises (or circular normal, see \cite{jammalamadaka2001topics}) distribution. This was confirmed using Watson’s test, where the test statistic (0.0259) was less than the critical value (0.079) at the 5\% significance level, leading to a failure to reject the null hypothesis.
Finally, Figure-\ref{simulated_plot_von_mises}(D) compares the exact regression curve (red line) with the fitted regression curve (blue line) over the simulated dataset, highlighting the accuracy of the  model.

\section{Comparison with some existing models}\label{comparison model sec}
 The proposed regression model shares some similarities with the models of \cite{fisher1992regression} and \cite{kim2015inverse}. \cite{fisher1992regression}  introduced a regression model where the link function is expressed as a form of the tangent function, while \cite{kim2015inverse} employed a specific stereographic projection as a link function. Both of these models rely on the von Mises distribution for the angular error. However, our model differs from theirs in several key aspects. First, it uses a general M\"{o}bius transformation that maps the real line to the unit circle. Second, the assumption about the angular error does not require to follow any particular parametric family of distribution, making it a semi-parametric regression model. Finally, we have developed a distinct cost function by applying tools from differential geometry to estimate the model parameters (constants from the M\"{o}biusmap).
 

\section{ Data Analysis} \label{data analysis sec}
\subsection{Data examples:} We are particularly interested in data associated with two popular cryptocurrencies: one is Bitcoin, and another is Ethereum. 
\begin{itemize}
    \item \textbf{Bitcoin:} Bitcoin, launched in 2009 by the pseudonymous Satoshi Nakamoto, operates on a decentralized peer-to-peer network using blockchain technology to enable secure, transparent transactions without intermediaries. With a capped supply of 21 million, Bitcoin is considered a deflationary asset and is often likened to digital gold. Its decentralized nature and scarcity make it a tool for transactions and wealth preservation, attracting global investors. The blockchain ensures transaction integrity, and the Proof of Work consensus mechanism secures the network. Bitcoin’s success has spurred the development of numerous other digital currencies, establishing it as a cornerstone in digital finance.

    \item  \textbf{Ethereum:} Introduced by Vitalik Buterin in 2015, Ethereum expanded blockchain technology beyond basic transactions, enabling the creation of smart contracts and decentralized applications (dApps). While Bitcoin primarily serves as a digital currency and store of value, Ethereum was designed for more versatile applications. Smart contracts on Ethereum automatically execute coded agreements without intermediaries, fostering trustless transactions. Ethereum’s native currency, Ether (ETH), powers these processes, incentivizing network participation. The Ethereum Virtual Machine (EVM) allows developers to build complex dApps, spurring innovation across banking, gaming, and supply chain sectors. Ongoing upgrades, including Ethereum 2.0 and a shift to Proof of Stake, aim to enhance scalability, security, and sustainability, solidifying Ethereum’s role as a cornerstone of blockchain technology.

\end{itemize}

\subsection{Data Processing:} We have collected per-minute historical datasets for Bitcoin and Ethereum, capturing high-frequency data to facilitate detailed analysis. Each dataset has columns for the Unix timestamp, date, symbol, opening price, highest price, lowest price, closing price, volume (in crypto), and volume (in the currency used as a base). These datasets provide a granular view of the price movements and associated timestamps, enabling precise modeling and exploration of trends, correlations, and angular relationships in financial time series data.

The date column denotes the timestamp in Coordinated Universal Time (UTC), the high column signifies the maximum price observed during the specified time, and the low column indicates the minimum price observed during the specified period. 

We focus on the data associated with the highest price. Since the data was recorded at per-minute intervals, each column contains 60 data points per hour. To reduce this dataset to a single observation per hour, we select the highest value recorded within each hour. This reduction results in 24 observations per day.
From these 24 hourly maximum values, we then identify the highest value across the day to determine a single observation representing the daily maximum.
To compute the angular value (in radians) corresponding to the timestamp of the daily maximum, we apply the following formula:
$$\theta= \frac{\displaystyle \arg \max_{1\leq j \leq 24}  \left(\max_{1\leq i \leq 60} v_i\right)_j}{24\times 60} \in [0,2\pi),$$ 
where $v_i$ is \textit{the highest price} of each minute. If we replace \textit{max }by \textit{min} in the above formula we will get the angular data corresponding to the \textit{lowest price}.\\

\begin{figure}[t!]
    \centering
    \subfloat[]{%
        {\includegraphics[trim= 2 2 2 2, clip,width=0.4\textwidth, height=0.4\textwidth]{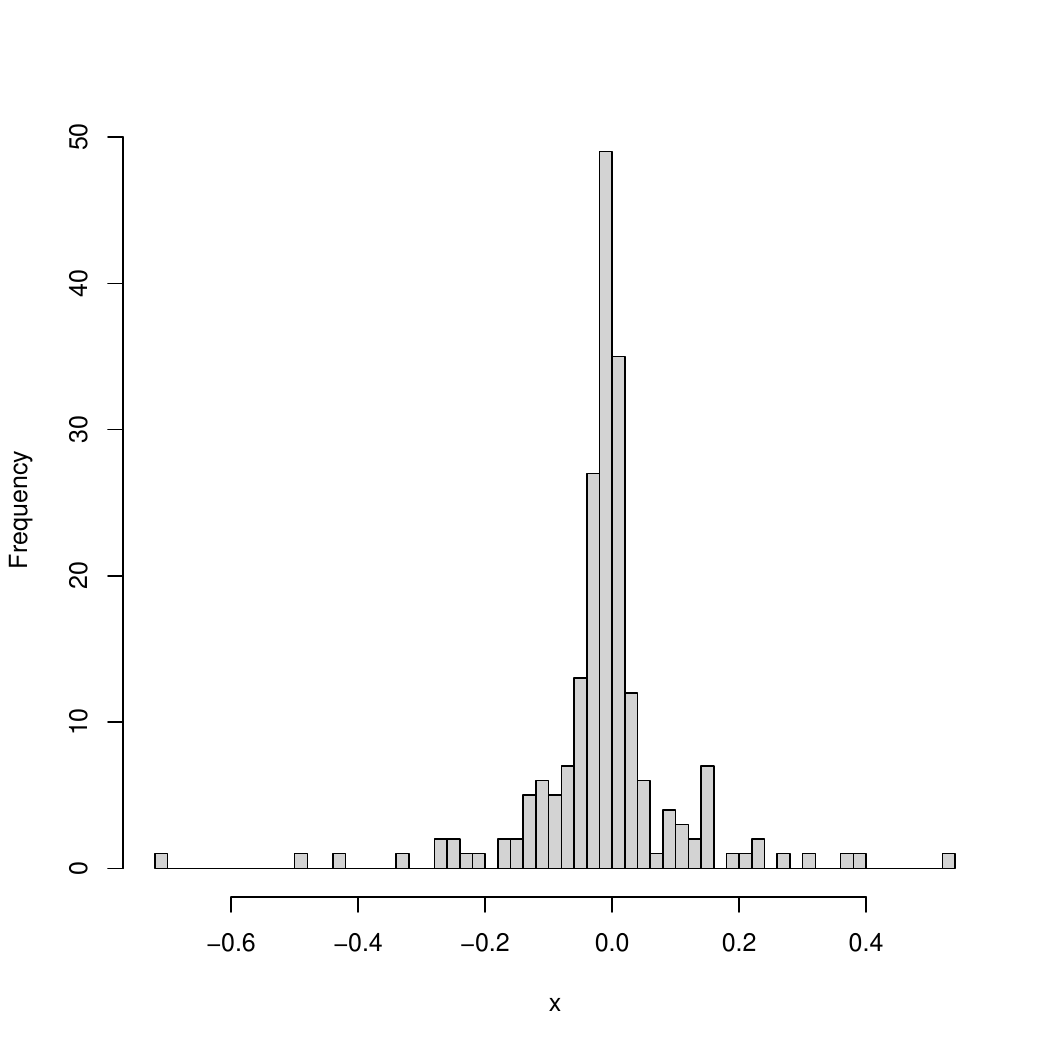}}}\hspace{5pt}
    \subfloat[ ]{%
        {\includegraphics[trim= 2 2 2 2, clip,width=0.4\textwidth, height=0.4\textwidth]{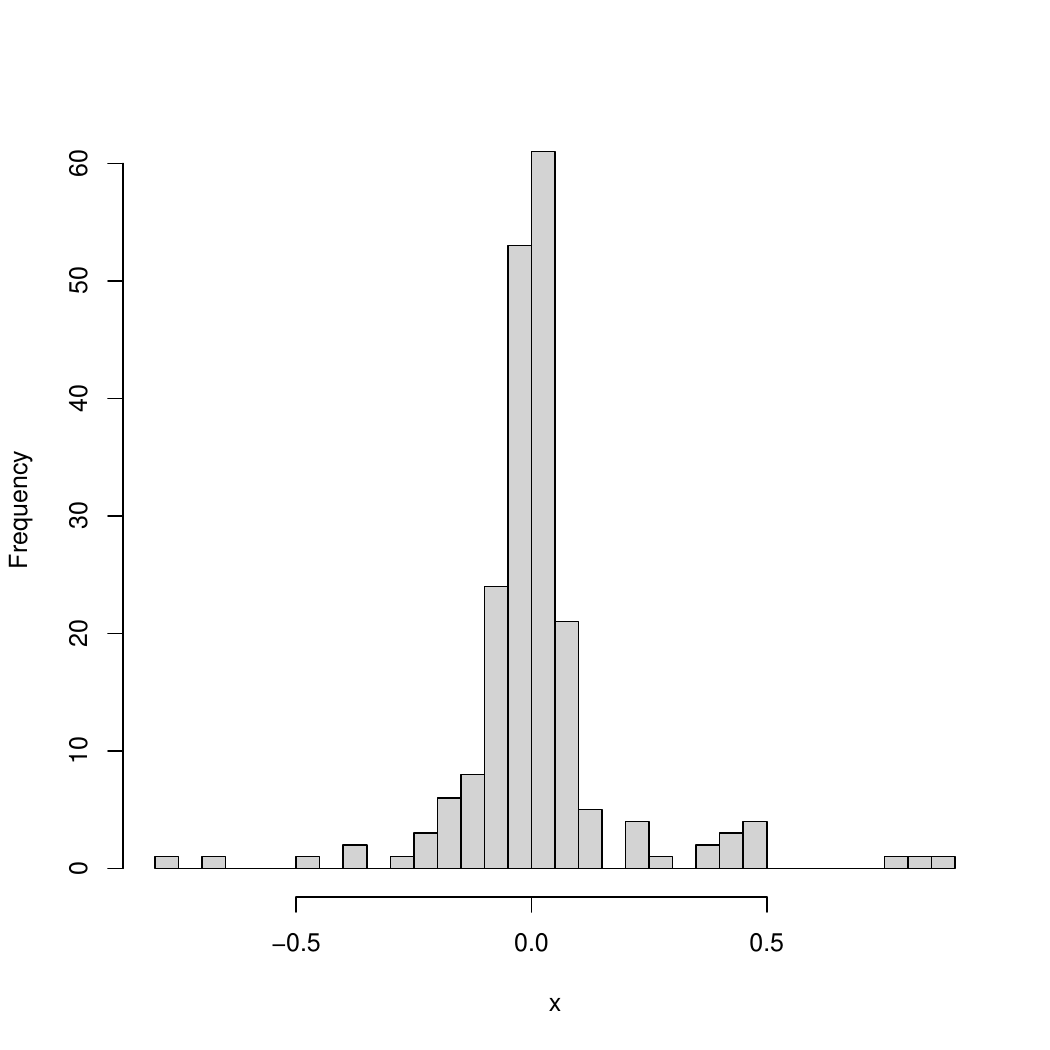}}}
   
    \subfloat[ ]{%
        {\includegraphics[trim= 2 2 2 2, clip,width=0.4\textwidth, height=0.4\textwidth]{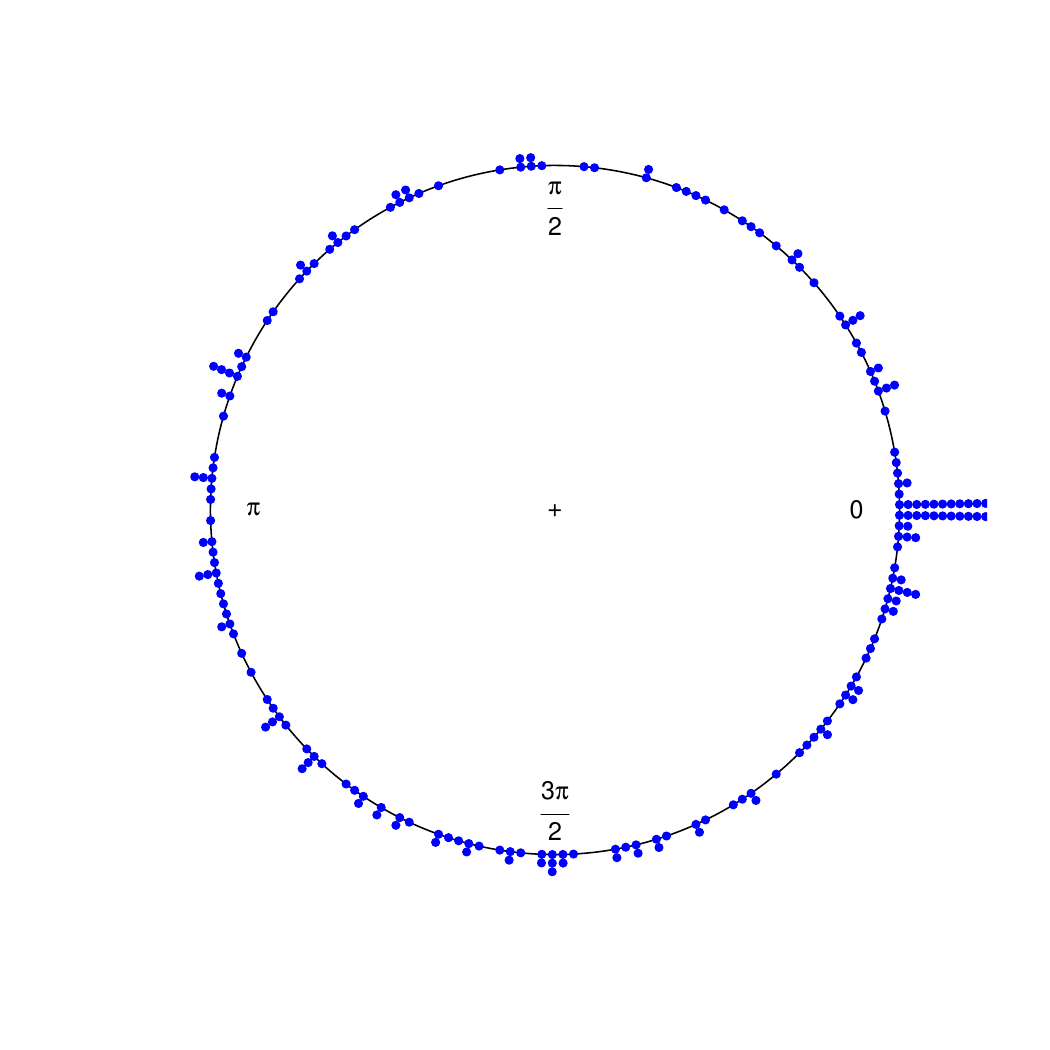}}}\vspace{5pt}
    \subfloat[ ]{%
        {\includegraphics[trim= 2 2 2 2, clip,width=0.4\textwidth, height=0.4\textwidth]{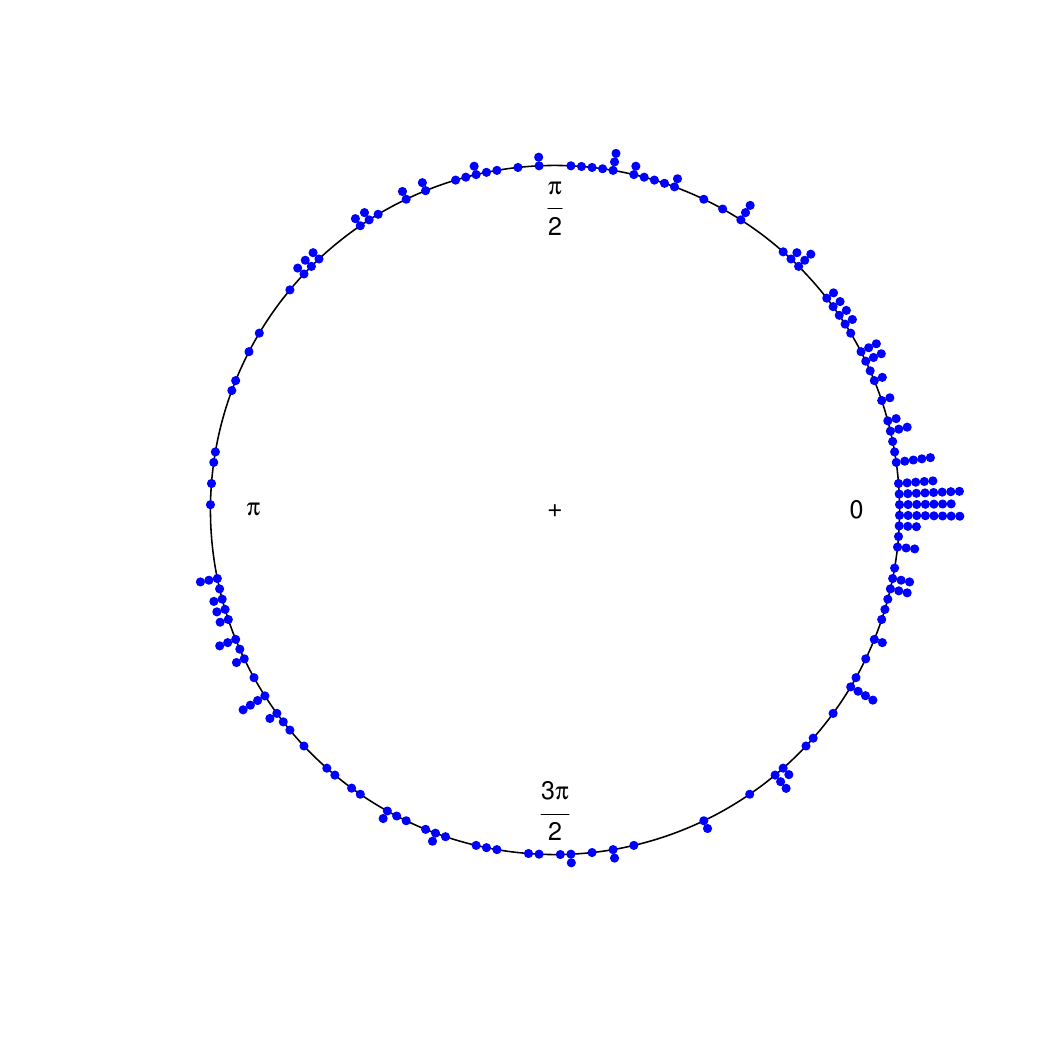}}}\vspace{5pt}
    \caption{(A) and (B): Histograms of the predictor for Bitcoin and Ethereum, respectively.  
(C) and (D): Circular plots of the response angles corresponding to the timestamps of the high prices for Bitcoin and Ethereum, respectively.}
     \label{hist_rose_bit_eth}
\end{figure}

\begin{figure}[t]
    \centering
    \subfloat[]{%
        {\includegraphics[trim= 2 2 2 2, clip,width=0.4\textwidth, height=0.4\textwidth]{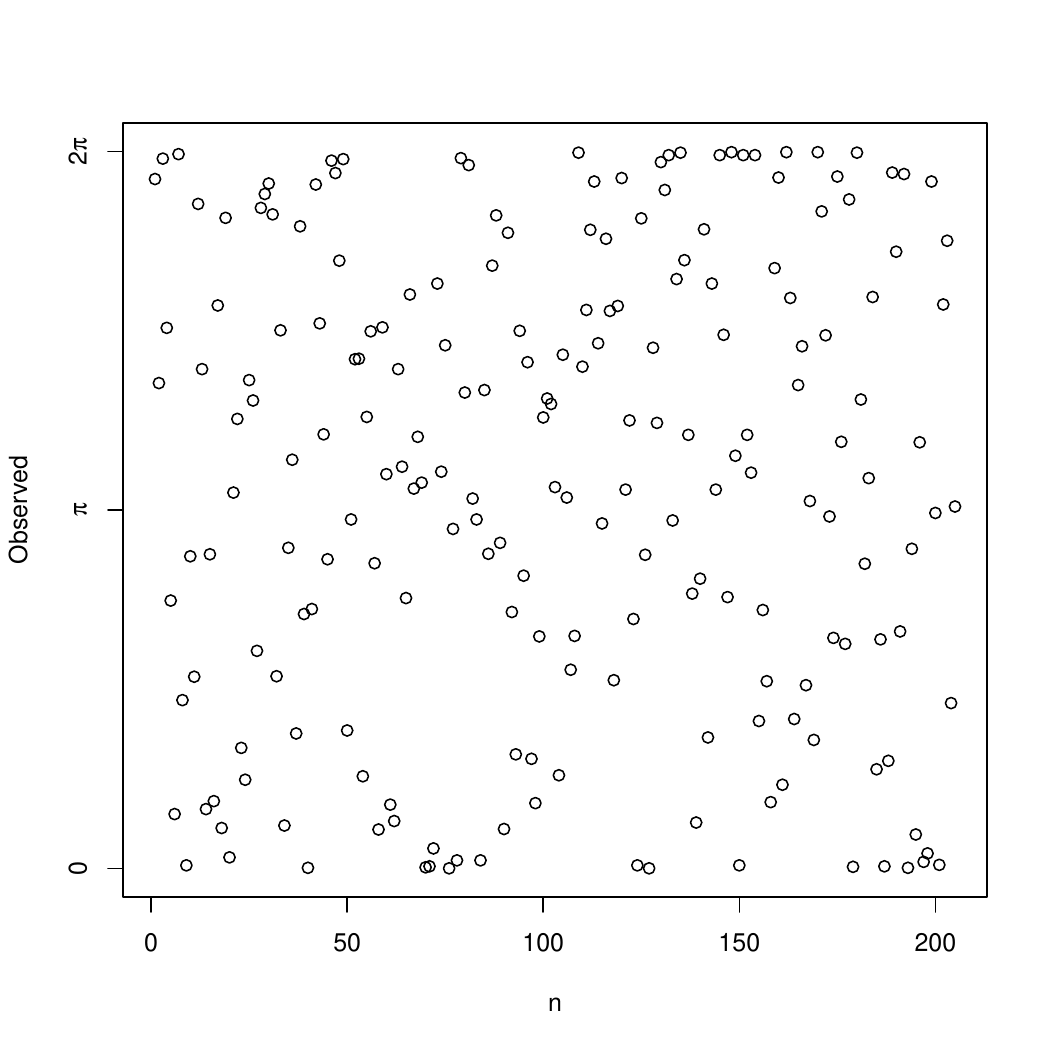}}}\hspace{5pt}
    \subfloat[ ]{%
        {\includegraphics[trim= 2 2 2 2, clip,width=0.4\textwidth, height=0.4\textwidth]{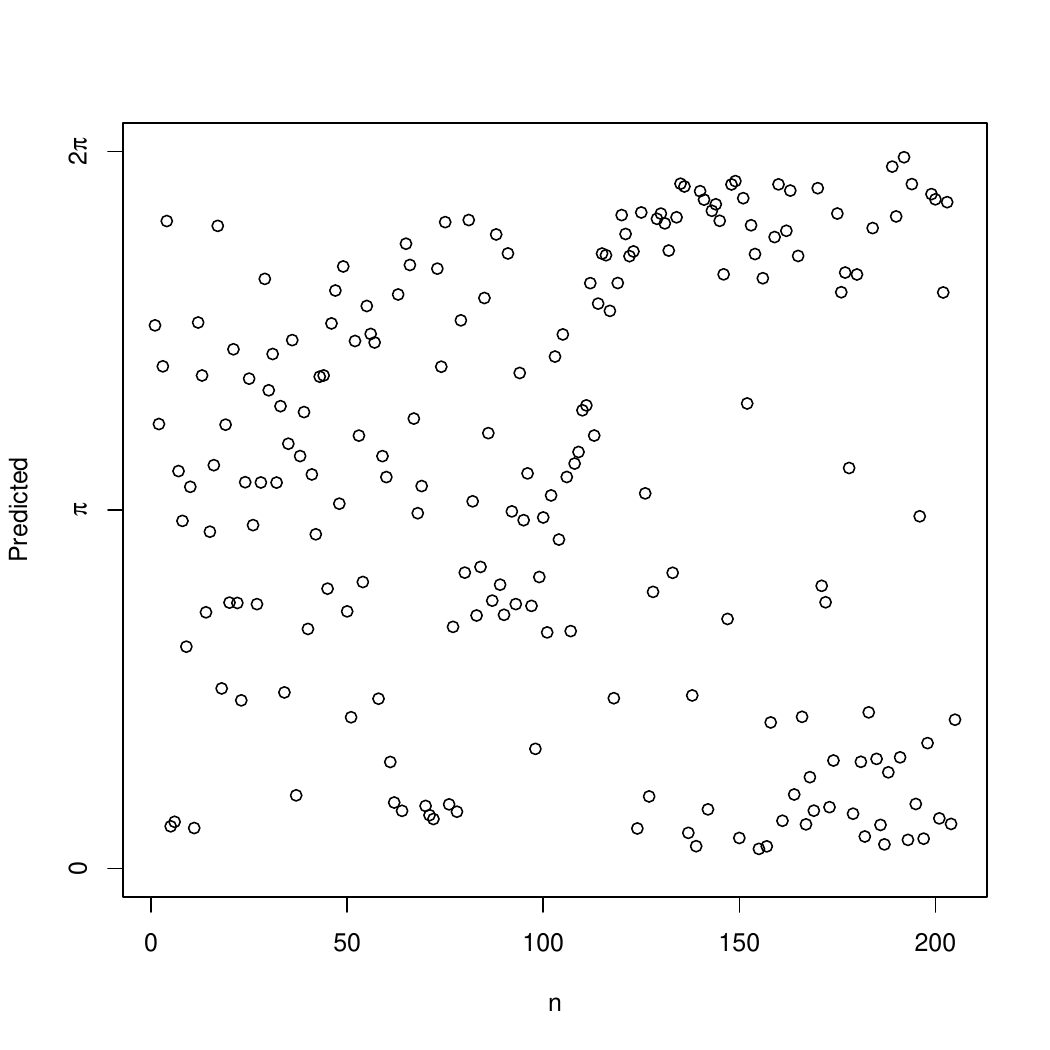}}}
   
    \subfloat[ ]{%
        {\includegraphics[trim= 2 2 2 2, clip,width=0.4\textwidth, height=0.4\textwidth]{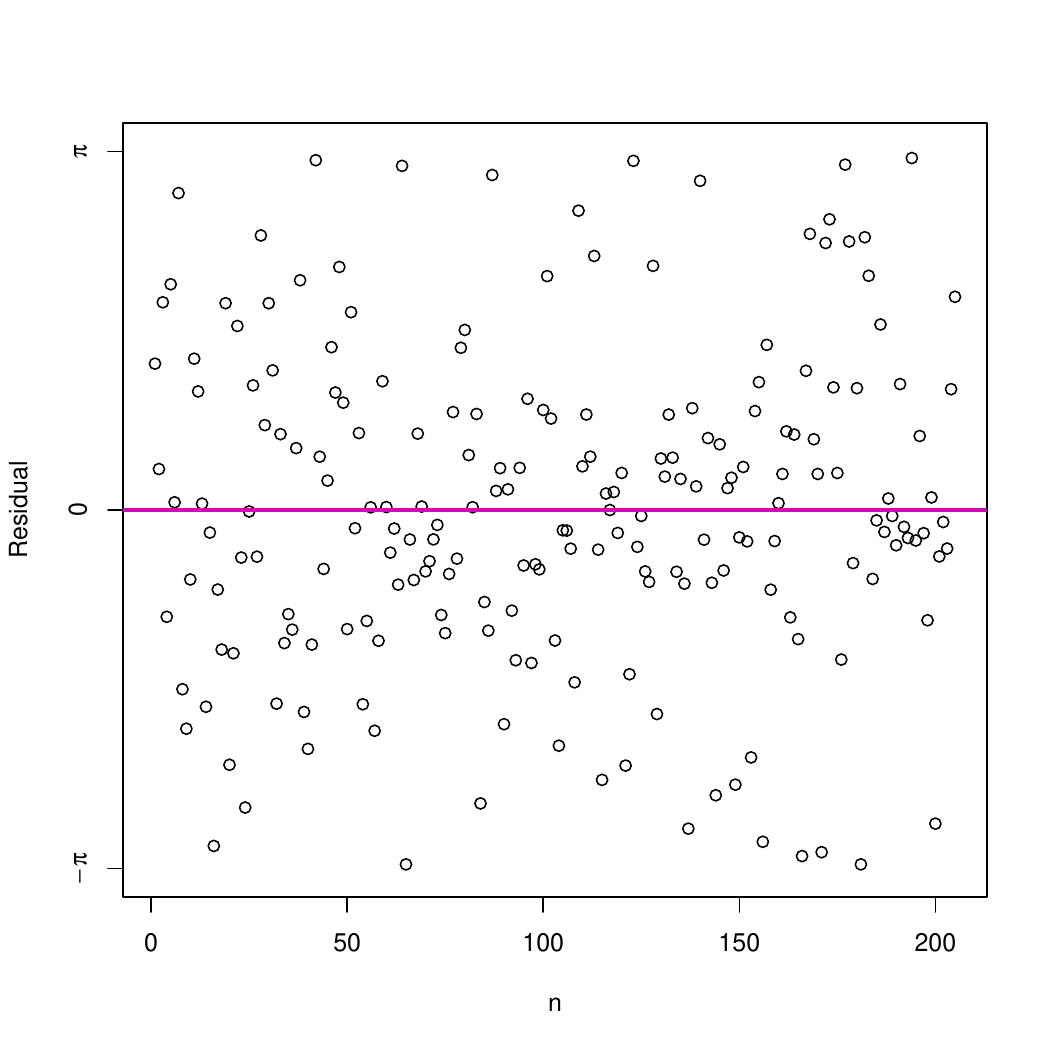}}}\vspace{5pt}
    \subfloat[ ]{%
        {\includegraphics[trim= 2 2 2 2, clip,width=0.4\textwidth, height=0.4\textwidth]{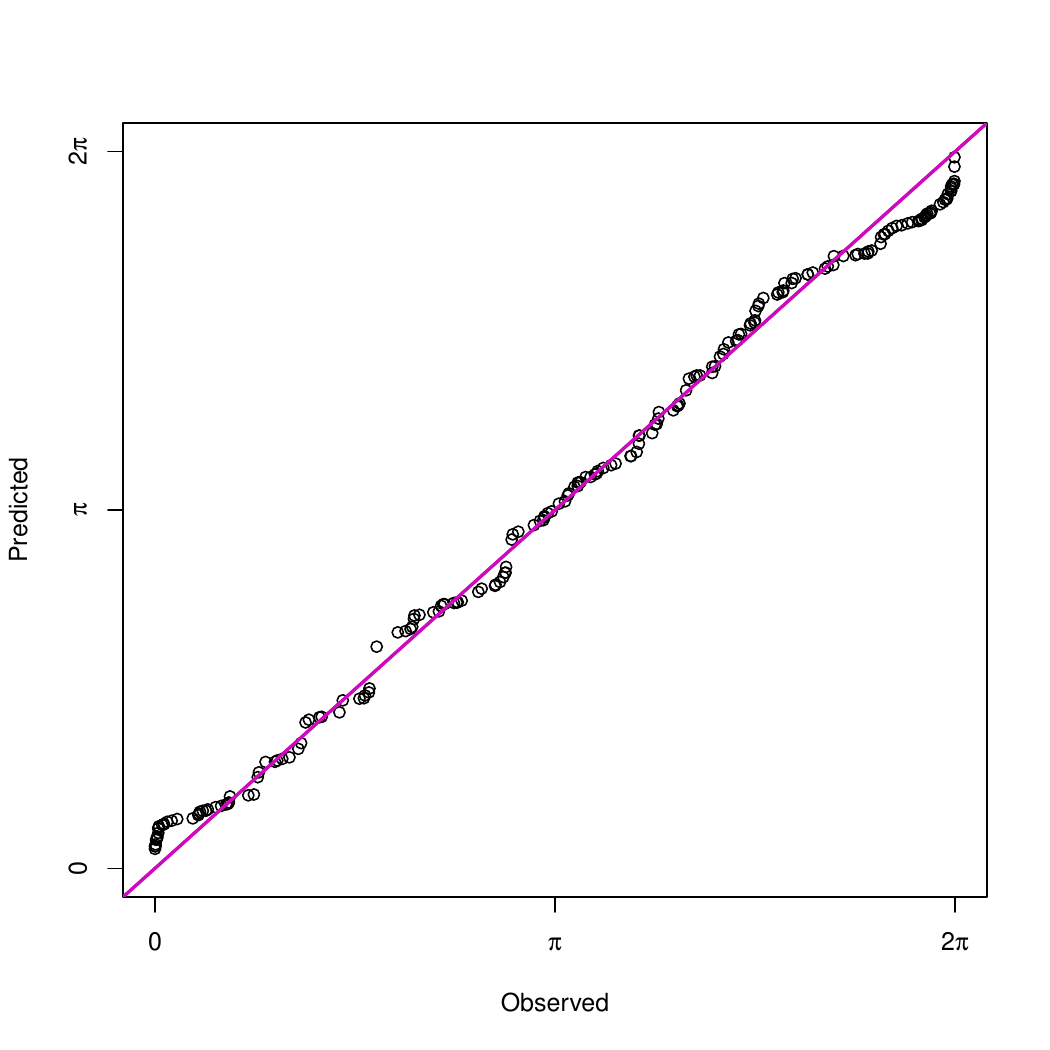}}}\vspace{5pt}
    \caption{ For Bitcoin: (A) Scatter plot of the angles corresponding to the timestamps of the high prices.  
(B) Scatter plot of the predicted angles for the timestamps of the high prices.  
(C) Plot of the residuals (restricted to the range \([-\pi, \pi]\) for improved visual clarity) with a reference line at Residual = 0. 
(D) QQ plot (on a radian scale) compares the observed angles with the predicted angles from the proposed model.}
     \label{data_plot_bit}
\end{figure}

\begin{figure}[t]
    \centering
    \subfloat[]{%
        {\includegraphics[trim= 2 2 2 2, clip,width=0.4\textwidth, height=0.4\textwidth]{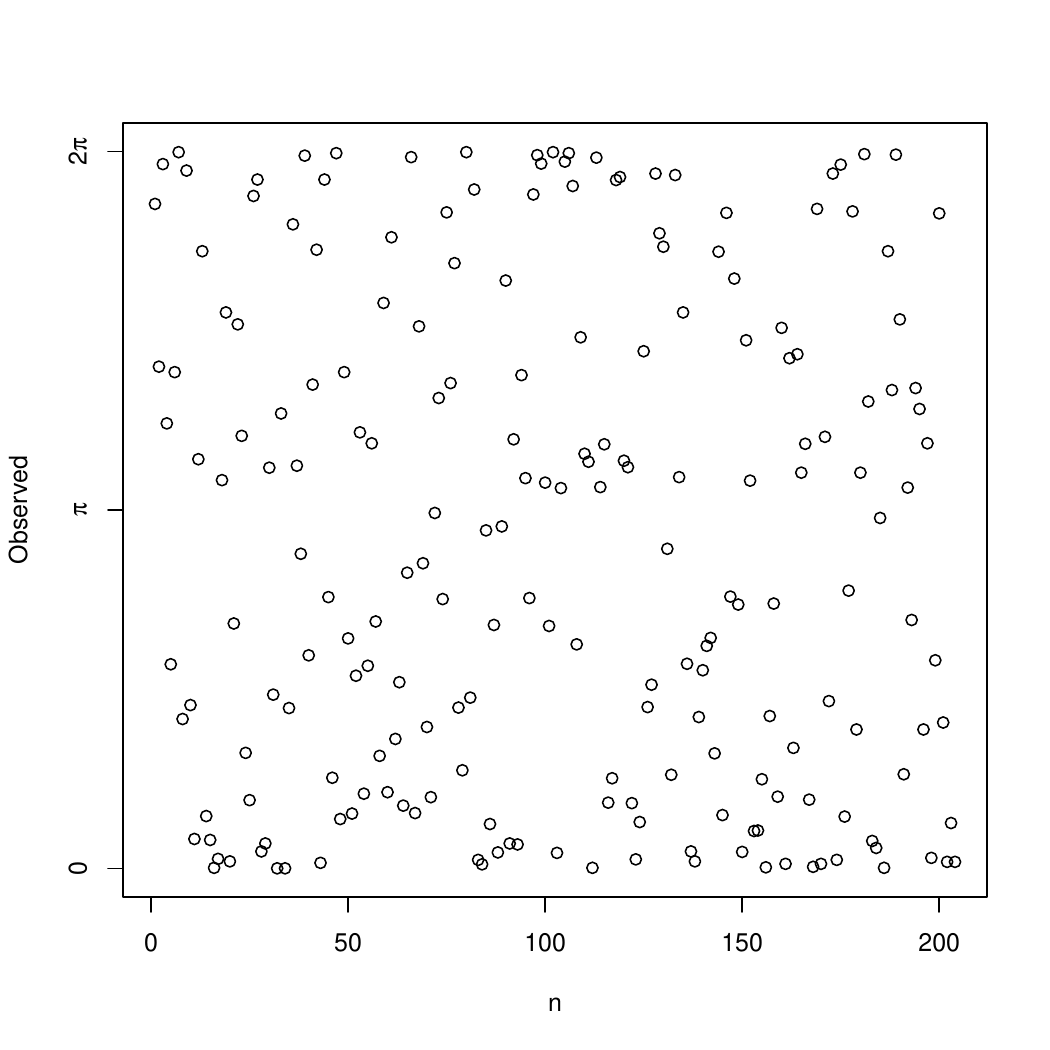}}}\hspace{5pt}
    \subfloat[ ]{%
        {\includegraphics[trim= 2 2 2 2, clip,width=0.4\textwidth, height=0.4\textwidth]{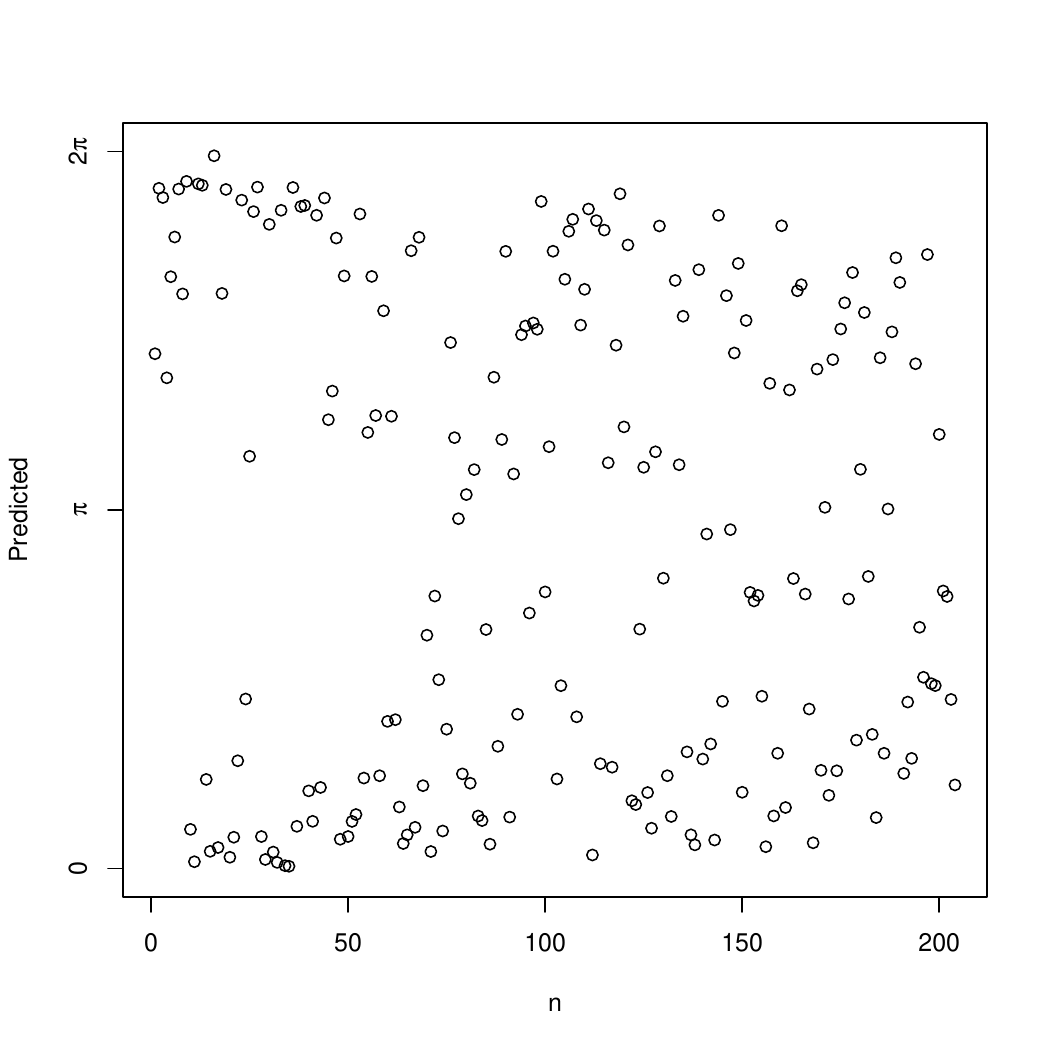}}}
   
    \subfloat[ ]{%
        {\includegraphics[trim= 2 2 2 2, clip,width=0.4\textwidth, height=0.4\textwidth]{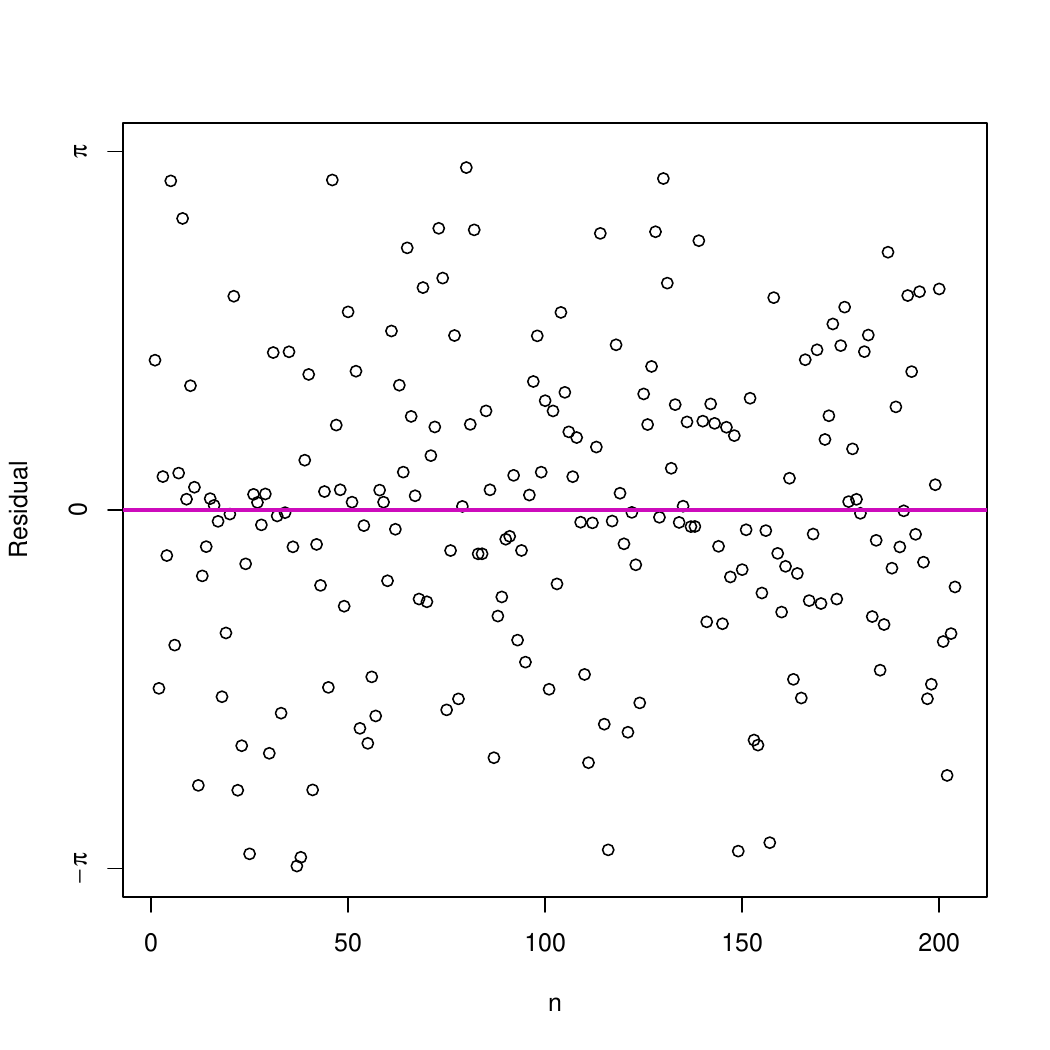}}}\vspace{5pt}
    \subfloat[ ]{%
        {\includegraphics[trim= 2 2 2 2, clip,width=0.4\textwidth, height=0.4\textwidth]{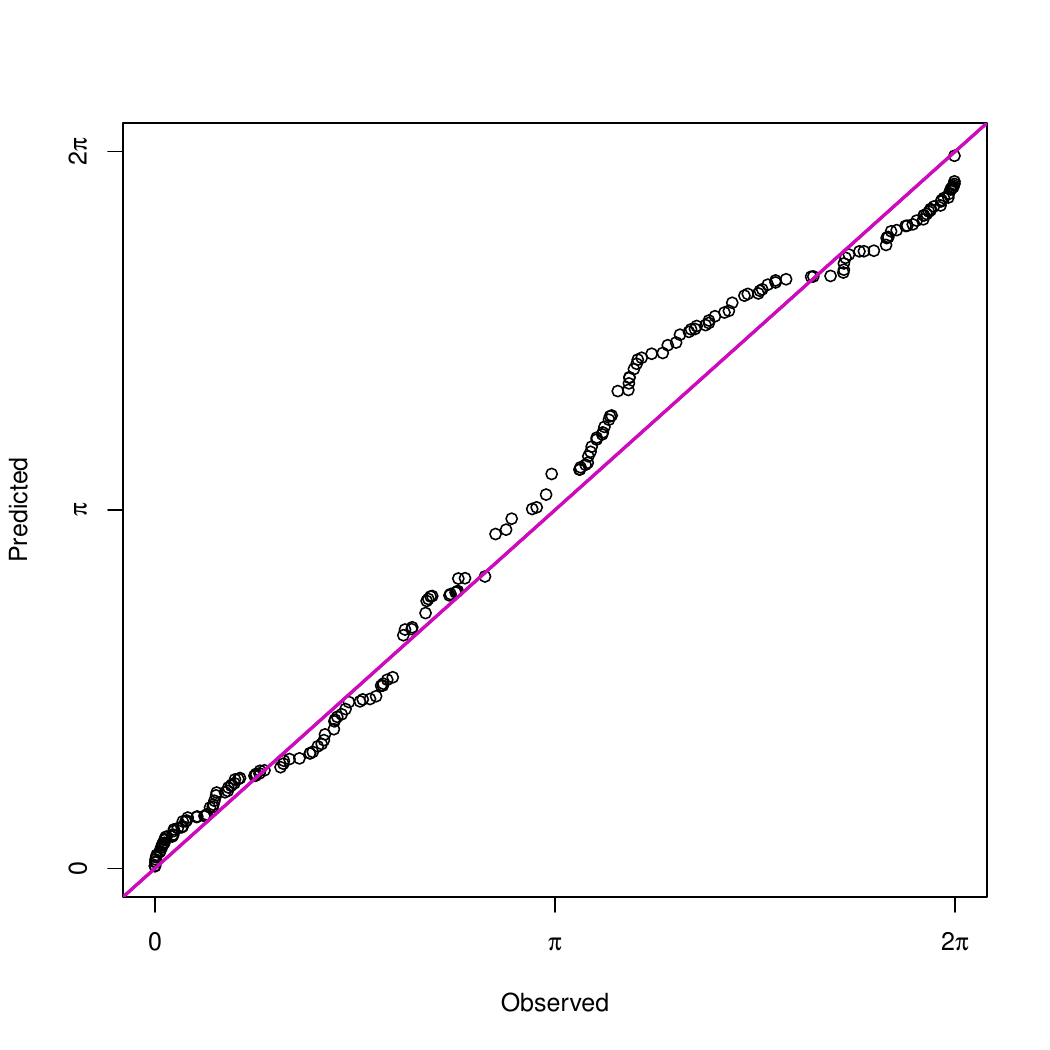}}}\vspace{5pt}
    \caption{For Ethereum:  
(A) Scatter plot of the angles corresponding to the timestamps of the high prices.  
(B) Scatter plot of the predicted angles for the timestamps of the high prices.  
(C) Plot of the residuals (restricted to the range \([-\pi, \pi]\) for improved visual clarity) with a reference line at Residual = 0.  
(D) QQ plot (on a radian scale) compares the observed angles with the predicted angles from the proposed model.}
     \label{data_plot_eth}
\end{figure}

\subsection{Analysis:}
In this section, we apply our proposed semi-parametric method to two popular cryptocurrencies: Bitcoin and Ethereum. For the proposed linear-circular regression technique, we collected a Bitcoin per-minute historical dataset from Kaggle (source: \url{https://www.kaggle.com/datasets/prasoonkottarathil/btcinusd?select=BTC-2017min.csv}). This \\dataset includes all one-minute historical data from January 1, 2017, to December 31, 2017, resulting in 365 daily observations after processing. However, this data is considered to be from January 1, 2017, to July 31, 2017, for the proposed regression model.
Similarly, we collected a per-minute historical dataset for Ethereum from Kaggle (source: \url{https://www.kaggle.com/datasets/prasoonkottarathil/ethereum-historical-dataset?select=ETH_1min.csv}) \\which covers the period from May 9, 2016, to April 16, 2020, yielding 1409 daily observations after processing. Here, we have considered the data from January 1, 2018, to July 31, 2018, for the proposed regression model.

In modeling the timestamp of extreme cryptocurrency values (a circular variable), the opening, closing, high, and low prices can be used as linear predictors, but their effective application requires addressing inherent complexities. Individually, these prices do not consistently correspond to specific times, as extreme values are influenced by factors such as sudden market news, liquidity shifts, and global trading activity. However, functions of these prices can encapsulate aspects like volatility, average trading levels, and directional trends that may correlate with temporal patterns. Employing such functional transformations makes it possible to reveal underlying relationships between price dynamics and the timing of extreme events, providing valuable insights into trading behavior and market structures. Specifically, we used the function  
\[
x = \frac{\text{lowest price}}{\text{highest price}} \times \frac{1}{\text{(closing price - opening price)}}
\]  
as a linear predictor, with the angular representation \(\theta\) of the timestamp of the high price serving as the circular response variable.

For this data analysis, we utilized the ``L-BFGS-B" numerical optimization method. The optimization process was performed 1,000 times, each with different initial values for the parameters. Specifically, \(b_0\) was initialized from a Uniform distribution \([0, 2\pi]\), \(b_1\) from a Uniform distribution \([-10, 10]\), and \(b_2\) from a Uniform distribution \([0, 10]\). The results presented in Table-\ref{table:data_ethereum_bit_table}
show the estimated parameters that achieved the minimum standard error and a reasonably good  QQ-plot that effectively captures the relationship in the data across these 1,000 iterations. 

For the Bitcoin, Figure-\ref{data_plot_bit}(A), and (B) show the true data and the predicted data, respectively. The residual plot in  Figure-\ref{data_plot_bit} (C) illustrates angular errors resulting from the proposed linear-circular regression model, with the x-axis representing the index of observations (\(n\)) and the y-axis showing residual values confined to the angular range $[-\pi,\pi)$. The residuals exhibit no visible trend or systematic pattern, suggesting the model captures the relationship between circular and linear components effectively. Most residuals cluster around zero, aligning with the assumption of minimal angular error under a well-fitted model. Additionally, the spread of residuals appears symmetric about the zero line. Analogous to residuals in linear regression that follow a normal distribution under well-met assumptions, the residuals in this model follow a von Mises distribution also known as the circular normal distribution. This is confirmed through Watson's test, where the test statistic $(0.0611)$ was less than the critical value $(0.079)$ at the $5\%$ significance level, leading to a failure to reject the null hypothesis for an upper tail test. Overall, the plot and statistical test results confirm the adequacy of the model, with the residuals being randomly distributed, centered around zero, and adhering to the theoretical assumptions for circular data. 
In the QQ-plot shown in Figure-\ref{data_plot_bit}(D), the horizontal axis represents the quantiles of the observed data, while the vertical axis represents the quantiles of the predicted values. The alignment of points along a line with a \(45^\circ\) slope indicates that the proposed linear-circular regression model effectively captures the relationship in the data. Similarly, in the case of Ethereum, Figure-\ref{data_plot_eth}(A), (B), (C), and (D) represent the true data, the predicted data, the residual plot, and the QQ-plot, respectively. Here, the residuals also follow a von Mises distribution, as confirmed by Watson's test. The test statistic of  $0.0581$  was below the critical value of $0.079 $ at the $5\%$ significance level, leading to a failure to reject the null hypothesis for an upper tail test.

\begin{table}[h!]
\centering
\renewcommand{\arraystretch}{1.4} 
\begin{tabular}{ |>{\centering\arraybackslash}p{2.5cm}|>{\centering\arraybackslash}p{2.5cm}|>{\centering\arraybackslash}p{2.5cm}|>{\centering\arraybackslash}p{2.5cm}|>{\centering\arraybackslash}p{2.5cm}| }
  \hline
Digital Currency & Estimated value of $b_0$ with standard error, $\hat{b}_0~(s.e)$ in radian & Estimated value of $b_1$ with standard error, $\hat{b}_1~(s.e)$ & Estimated value of $b_2$ with standard error, $\hat{b}_2~(s.e)$ & Correlation \\ 
  \hline\hline
  Bitcoin &  2.7836 (4.8660) & -0.0068(0.1511) &  0.0362 (0.3097)& 0.7430  \\ 
  \hline
  Ethereum &  2.2596 (4.5086) & -0.0310 (0.3548) &  0.0702  (0.5870)& 0.7597\\ 
  \hline
\end{tabular}
\vspace{0.3cm}
\caption{Estimated parameter values (with standard errors) and correlations between the predictor and response for Bitcoin and Ethereum data.}
\label{table:data_ethereum_bit_table}
\end{table}


\section{ Conclusion} \label{conclusion}
In this paper, we have introduced a novel area-based loss function for the regression model formulated for the angular response variable where the predictor is linear. By leveraging the generalized M\"{o}bius transformation to define the regression curve, we seamlessly map the real axis to the circle, capturing the intrinsic relationship between linear and angular components. A key feature of our model is the area-based loss function, rooted in the geometry of a curved torus, which provides an efficient estimate of parameters. Importantly, the semi-parametric nature of the model eliminates the need for specific distributional assumptions about the angular error, widening its applicability. With a
through extensive simulation study using von Mises and wrapped Cauchy distributions, we have demonstrated the robustness and flexibility of the proposed framework. The practical utility of the model was further validated through real-world data analysis involving two major cryptocurrencies, Bitcoin and Ethereum. It shows the capability of the model to extract meaningful insights from complex features of the data. These results highlight the potential of the proposed regression model as a powerful tool for analyzing linear-circular relationships in various applied contexts.

\section{Acknowledgement} The authors are thankful to Mr. Shabhunath Sen, a doctoral candidate at the Department of Mathematics, Indian Institute of Technology Kharagpur, India, for helpful discussion.

\section{Funding} S. Biswas expresses gratitude for the financial support received through a Junior/Senior Research Fellowship from the Ministry of Human Resource Development (MHRD) and IIT Kharagpur, India. B. Banerjee acknowledges the funding provided by the Science and Engineering Research Board (SERB), Department of Science and Technology, Government of India, under the MATRICS grant (File No. MTR/2021/000397).


\bibliographystyle{apalike}
\bibliography{buddha_bib}
\newpage




    



\end{document}